




%

%
 \font\twelvebf=cmbx12
 \font\twelvett=cmtt12
 \font\twelveit=cmti12
 \font\twelvesl=cmsl12
 \font\twelverm=cmr12		\font\ninerm=cmr9
 \font\twelvei=cmmi12		\font\ninei=cmmi9
 \font\twelvesy=cmsy10 at 12pt	\font\ninesy=cmsy9
 \skewchar\twelvei='177		\skewchar\ninei='177
 \skewchar\seveni='177	 	\skewchar\fivei='177
 \skewchar\twelvesy='60		\skewchar\ninesy='60
 \skewchar\sevensy='60		\skewchar\fivesy='60
%
%

%
 \font\fourteenrm=cmr12 scaled 1200
 \font\seventeenrm=cmr12 scaled 1440
 \font\fourteenbf=cmbx12 scaled 1200
 \font\seventeenbf=cmbx12 scaled 1440
%
%

%
\batchmode
\font\tenmsy=msym10

\font\tenmsb=msbm10
\font\twelvemsb=msbm10 scaled 1200
\errorstopmode

\edef\ff{\fontname\tenmsy}
\def\m{m}
\edef\fa{\if\m\ff\fi}
\def\m{s}
\edef\fa{\if\m\fa\fi}
\def\m{y}
\edef\fa{\if\m\fa\fi}
\def\m{m}
\edef\fa{\if\m\fa\fi}
\def\m{1}
\edef\fa{\if\m\fa\fi}
\def\m{0}
\if\m\fa

\let\twelvebkb=\twelvemsy
\else

\let\twelvebkb=\twelvemsb
\fi
\newfam\msbfam

%
\font\tensc=cmcsc10
\font\twelvesc=cmcsc10 scaled 1200
\newfam\scfam

%
\def\seventeenpt{\def\rm{\fam0\seventeenrm}%
 \textfont\bffam=\seventeenbf	\def\bf{\fam\bffam\seventeenbf}}
\def\fourteenpt{\def\rm{\fam0\fourteenrm}%
 \textfont\bffam=\fourteenbf	\def\bf{\fam\bffam\fourteenbf}}
\def\twelvept{\def\rm{\fam0\twelverm}%
 \textfont0=\twelverm	\scriptfont0=\ninerm	\scriptscriptfont0=\sevenrm
 \textfont1=\twelvei	\scriptfont1=\ninei	\scriptscriptfont1=\seveni
 \textfont2=\twelvesy	\scriptfont2=\ninesy	\scriptscriptfont2=\sevensy
 \textfont3=\tenex	\scriptfont3=\tenex	\scriptscriptfont3=\tenex
 \textfont\itfam=\twelveit	\def\it{\fam\itfam\twelveit}%
 \textfont\slfam=\twelvesl	\def\sl{\fam\slfam\twelvesl}%
 \textfont\ttfam=\twelvett	\def\tt{\fam\ttfam\twelvett}%
 \scriptfont\bffam=\tenbf 	\scriptscriptfont\bffam=\sevenbf
 \textfont\bffam=\twelvebf	\def\bf{\fam\bffam\twelvebf}%
 \textfont\scfam=\twelvesc	\def\sc{\fam\scfam\twelvesc}%
 \textfont\msbfam=\twelvemsb	
 \baselineskip 14pt%
 \abovedisplayskip 7pt plus 3pt minus 1pt%
 \belowdisplayskip 7pt plus 3pt minus 1pt%
 \abovedisplayshortskip 0pt plus 3pt%
 \belowdisplayshortskip 4pt plus 3pt minus 1pt%
 \parskip 3pt plus 1.5pt
 \setbox\strutbox=\hbox{\vrule height 10pt depth 4pt width 0pt}}
\def\tenpt{\def\rm{\fam0\tenrm}%
 \textfont0=\tenrm	\scriptfont0=\sevenrm	\scriptscriptfont0=\fiverm
 \textfont1=\teni	\scriptfont1=\seveni	\scriptscriptfont1=\fivei
 \textfont2=\tensy	\scriptfont2=\sevensy	\scriptscriptfont2=\fivesy
 \textfont3=\tenex	\scriptfont3=\tenex	\scriptscriptfont3=\tenex
 \textfont\itfam=\tenit		\def\it{\fam\itfam\tenit}%
 \textfont\slfam=\tensl		\def\sl{\fam\slfam\tensl}%
 \textfont\ttfam=\tentt		\def\tt{\fam\ttfam\tentt}%
 \scriptfont\bffam=\sevenbf 	\scriptscriptfont\bffam=\fivebf
 \textfont\bffam=\tenbf		\def\bf{\fam\bffam\tenbf}%
 \textfont\scfam=\tensc		\def\sc{\fam\scfam\tensc}%
 \textfont\msbfam=\tenmsb	
 \baselineskip 12pt%
 \abovedisplayskip 6pt plus 3pt minus 1pt%
 \belowdisplayskip 6pt plus 3pt minus 1pt%
 \abovedisplayshortskip 0pt plus 3pt%
 \belowdisplayshortskip 4pt plus 3pt minus 1pt%
 \parskip 2pt plus 1pt
 \setbox\strutbox=\hbox{\vrule height 8.5pt depth 3.5pt width 0pt}}

%
\def\twelvepoint{%
 \def\small{\tenpt\rm}%
 \def\normal{\twelvept\rm}%
 \def\big{\fourteenpt\rm}%
 \def\huge{\seventeenpt\rm}%
 \footline{\hss\twelverm\folio\hss}
 \normal}
\def\tenpoint{%
 \def\small{\tenpt\rm}%
 \def\normal{\tenpt\rm}%
 \def\big{\twelvept\rm}%
 \def\huge{\fourteenpt\rm}%
 \footline{\hss\tenrm\folio\hss}
 \normal}

%
\def\bigbold{\big\bf}

%
\catcode`\@=11
%
%
\def\footnote#1{\edef\@sf{\spacefactor\the\spacefactor}#1\@sf
 \insert\footins\bgroup\small
 \interlinepenalty100	\let\par=\endgraf
 \leftskip=0pt		\rightskip=0pt
 \splittopskip=10pt plus 1pt minus 1pt	\floatingpenalty=20000
 \smallskip\item{#1}\bgroup\strut\aftergroup\@foot\let\next}
%
%
%
%
\def\hexnumber@#1{\ifcase#1 0\or 1\or 2\or 3\or 4\or 5\or 6\or 7\or 8\or
 9\or A\or B\or C\or D\or E\or F\fi}
\edef\msbfam@{\hexnumber@\msbfam}

%
%
%
\catcode`\@=12

\newcount\EQNO      \EQNO=0
\newcount\FIGNO     \FIGNO=0
\newcount\REFNO     \REFNO=0
\newcount\SECNO     \SECNO=0
\newcount\SUBSECNO  \SUBSECNO=0
\newcount\FOOTNO    \FOOTNO=0
\newbox\FIGBOX      \setbox\FIGBOX=\vbox{}
\newbox\REFBOX      \setbox\REFBOX=\vbox{}
\newbox\RefBoxOne   \setbox\RefBoxOne=\vbox{}

\expandafter\ifx\csname normal\endcsname\relax\def\normal{\null}\fi

\def\Eqno{\global\advance\EQNO by 1 \eqno(\the\EQNO)%
    \gdef\label##1{\xdef##1{\nobreak(\the\EQNO)}}}
\def\Fig#1{\global\advance\FIGNO by 1 Figure~\the\FIGNO%
    \global\setbox\FIGBOX=\vbox{\unvcopy\FIGBOX
      \narrower\smallskip\item{\bf Figure \the\FIGNO~~}#1}}
\def\Ref#1{\global\advance\REFNO by 1 \nobreak[\the\REFNO]%
    \global\setbox\REFBOX=\vbox{\unvcopy\REFBOX\normal
      \smallskip\item{\the\REFNO .~}#1}%
    \gdef\label##1{\xdef##1{\nobreak[\the\REFNO]}}}
\def\Section#1{\SUBSECNO=0\advance\SECNO by 1
    \bigskip\leftline{\bf \the\SECNO .\ #1}\nobreak}
\def\Subsection#1{\advance\SUBSECNO by 1
    \medskip\leftline{\bf \ifcase\SUBSECNO\or
    a\or b\or c\or d\or e\or f\or g\or h\or i\or j\or k\or l\or m\or n\fi
    )\ #1}\nobreak}
\def\Footnote#1{\global\advance\FOOTNO by 1
    \footnote{\nobreak$\>\!{}^{\the\FOOTNO}\>\!$}{#1}}
\def\SameFootnote{$\>\!{}^{\the\FOOTNO}\>\!$}

\def\References{\bigskip\centerline{\bf REFERENCES}
                \smallskip\copy\REFBOX}
\def\NewRefPage{\setbox\RefBoxOne=\vbox{\unvcopy\REFBOX}
		\setbox\REFBOX=\vbox{}
		\def\References{\bigskip\centerline{\bf REFERENCES}
                		\nobreak\smallskip\nobreak\copy\RefBoxOne
				\vfill\eject
				\smallskip\copy\REFBOX}
		\def\NewRefPage{}}




\font\twelvebm=cmmib10 at 12pt
\font\tenbm=cmmib10
\font\ninei=cmmi9
\newfam\bmfam

\def\twelvepointbmit{
\textfont\bmfam=\twelvebm
\scriptfont\bmfam=\ninei
\scriptscriptfont\bmfam=\seveni
\def\bmit{\fam\bmfam\twelvebm}
}

\def\tenpointbmit{
\textfont\bmfam=\tenbm
\scriptfont\bmfam=\seveni
\scriptscriptfont\bmfam=\fivei
\def\bmit{\fam\bmfam\tenbm}
}

\tenpointbmit

\mathchardef\Gamma="7100
\mathchardef\Delta="7101
\mathchardef\Theta="7102
\mathchardef\Lambda="7103
\mathchardef\Xi="7104
\mathchardef\Pi="7105
\mathchardef\Sigma="7106
\mathchardef\Upsilon="7107
\mathchardef\Phi="7108
\mathchardef\Psi="7109
\mathchardef\Omega="710A
\mathchardef\alpha="710B
\mathchardef\beta="710C
\mathchardef\gamma="710D
\mathchardef\delta="710E
\mathchardef\epsilon="710F
\mathchardef\zeta="7110
\mathchardef\eta="7111
\mathchardef\theta="7112
\mathchardef\iota="7113
\mathchardef\kappa="7114
\mathchardef\lambda="7115
\mathchardef\mu="7116
\mathchardef\nu="7117
\mathchardef\xi="7118
\mathchardef\pi="7119
\mathchardef\rho="711A
\mathchardef\sigma="711B
\mathchardef\tau="711C
\mathchardef\upsilon="711D
\mathchardef\phi="711E
\mathchardef\cho="711F
\mathchardef\psi="7120
\mathchardef\omega="7121
\mathchardef\varepsilon="7122
\mathchardef\vartheta="7123
\mathchardef\varpi="7124
\mathchardef\varrho="7125
\mathchardef\varsigma="7126
\mathchardef\varphi="7127



\twelvepoint
\magnification 1000

\parskip 1em
\parindent 0em
\def\qed{{\vrule height 6pt width 6pt}}
\def\bp{{\rm\, B\! p}}
\def\hh{\widehat}

\def\S{{\rm\bf S }}

\def\E{{\rm\bf E }}
\def\L{{\rm\bf L }}
\def\M{{\rm\bf M }}
\def\LuM{{\L^\al\union\M^\al}}
\def\LuMbe{{\L^\be\union\M^\be}}
\def\EuS{{\E^\al\union\S^\al}}

\def\Ual{{\rm\bf U}^\al}
\def\Ube{{\rm\bf U}^\be}

\def\Eal{{\rm\bf E}^\al}
\def\Lal{{\rm\bf L}^\al}
\def\Mal{{\rm\bf M}^\al}
\def\ME{{M^\E}}
\def\Sigal{\Sigma^\al}
\def\Sigbe{\Sigma^\be}
\def\Phil{\Phi^\al}
\def\Phibe{\Phi^\be}

\def\al{{\scriptstyle \alpha}}
\def\be{{\scriptstyle \beta}}

\def\del{{\delta}}
\def\lal{{\alpha}}
\def\lbe{{\beta}}
\def\albe{{(\!\alpha\beta\!)}}
\def\beal{{(\!\beta\alpha\!)}}

\def\tauSig{\tau_{\lower1pt\hbox{$\scriptscriptstyle\bf\Sigma$}}}
\def\tauM{\tau_{\lower1pt\hbox{$\scriptscriptstyle\bf M$}}}
\def\tauS{\tau_{\lower1pt\hbox{$\scriptscriptstyle\bf S$}}}


\def\projal{{p^\al}}
\def\projbe{{p^\be}}
\def\prohal{{\hh p^\al}}
\def\prohbe{{\hh p^\be}}
\def\prohn#1{{\hh p^{(#1)}}}
\def\Proj{{\cal P}}
\def\tmone{{}^{-1}}
\def\aabn#1{{a^\albe_#1}}

\def\cenal{{s_\alpha}}
\def\oppal{{s^\infty_\alpha}}
\def\cenbe{{s_\beta}}
\def\oppbe{{s^\infty_\beta}}

\def\cendel{{s_\delta}}

\def\alinf{{\al^\infty}}
\def\beinf{{\be^\infty}}
\def\Mob{{\cal M}}

\def\Cap{{\rm Cap}}
\def\cylinder{{\rm cylinder}}

\def\Rsq#1{{R_#1 ^{\;2}}}

\def\Intg{{\hbox{\twelvebkb Z} }}
\def\Real{{\hbox{\twelvebkb R} }}
\def\Cmpx{{\hbox{\twelvebkb C} }}
\def\Cminf{{\hbox{\twelvebkb C}_\infty }}
\def\union{\cup}
\def\Union{\bigcup}
\def\inter{\cap}
\def\Re{{\rm Re}}
\def\Im{{\rm Im}}

\def\Gt{G(\tau)}
\def\Tau{{\rm T}}
\def\lie{{\rm L}}
\def\ang{{\cal L}}
\def\eng{{\cal E}}
\def\Intr{{\cal I}}
\def\Ang{{\hbox{\twelvebkb L}}}
\def\cnj{\overline}

\def\dbydtau{{\partial\over\partial\tau}}
\def\dbydphi{{\partial\over\partial\phi}}

\def\newline{\hfil\break}
\def\half{{\textstyle{1 \over 2}}}
\def\ibytwo{{\textstyle{i \over 2}}}

\def\omegabytwo{{\scriptstyle\left({\omega \over 2}\right)}}

\def\oneoverfourR{{\textstyle{1 \over 4R_1}}}


%
%
%
%
\twelvepointbmit		

\def\sqr#1#2{{\vbox{\hrule height.#2pt
 	\hbox{\vrule width.#2pt height#1pt \kern#1pt\vrule width.#2pt}
		\hrule height.#2pt}}}

\def\Partial#1{\partial_{#1}^{\raise2pt\hbox{$\scriptstyle 2$}}}

\def\Uin{u^{\,\raise2pt\hbox{$\scriptstyle\rm in$}}}
\def\Uout{u^{\,\raise2pt\hbox{$\scriptstyle\rm out$}}}


\nopagenumbers

\def\today{\number\day\space\ifcase\month\or
  January\or February\or March\or April\or May\or June\or
  July\or August\or September\or October\or November\or December\fi
  \space\number\year}
\bigskip\bigskip

\null\bigskip\bigskip\bigskip

\baselineskip=27pt

\centerline{\bigbold   Topology Change  and the Propagation of
Massless Fields}
\bigskip\bigskip\bigskip

\centerline{Jonathan Gratus}
\centerline{Robin W Tucker}

\medskip

\centerline{\it School of Physics and Materials,}
\centerline{\it University of Lancaster,
		Bailrigg, Lancs. LA1 4YB, UK}
\centerline{\tt rwt{\rm @}lavu.physics.lancaster.ac.uk}

\bigskip\bigskip\bigskip\bigskip
\midinsert
\narrower\narrower\noindent


We analyse the massless wave equation on a class of
 two dimensional manifolds consisting of an arbitrary number of topological
cylinders connected to one or more topological spheres. Such manifolds are
endowed  with a degenerate (non-globally hyperbolic) metric.
Attention is drawn to the topological constraints on solutions describing
monochromatic modes on both compact and non-compact manifolds.
Energy and momentum currents are constructed and a new global sum rule
discussed. The results offer a rigorous background for the formulation of a
field theory of topologically induced particle production.

\endinsert

\vfill

\eject

\headline={\hss\rm -~\folio~- \hss}     

\def\frac#1#2{{#1\over #2}}

\Section{\bf  Introduction}


It is well known that the global structure of a manifold is fundamental in
constructing regular  solutions to tensor equations.
Furthermore field quantisation is sensitive to the  topology of the underlying
base space
 \Ref{S J Avis, C J Isham, Recent Developments in Gravitation, Ed. M L\`evy
, S Deser, Carg\`ese, Plenum Press, 1978},  \Ref{A Ashketar, {\it Lectures on
Non-Perturbative Canonical Gravity}, Advanced series in Astrophysics and
Cosmology, Vol 6, World Scientific, 1991}, \Ref{ J S Dowker, J Phys. A
{\bf 10} (1967) 115, {\bf 5} (1971) 1375 }.
However little attention has been given to the elucidation
of classical solutions of field equations on manifolds with degenerate
geometries that can accommodate non-trivial  topology
change in general relativity. Such solutions arise from equations that  are not
globally
hyperbolic.

Some of the earliest mathematical work on the study of partial
differential equations that change from being hyperbolic to elliptic was
done by Tricomi
 \Ref{M M Smirnov,  Trans. Amer. Math. Monographs (Amer. Math. Soc) {\bf 51}
1957
}.
This early treatise involved
considerable technicalities that have not been extensively pursued in the
mathematical literature.
Even in two dimensions
the analysis of second order
 partial differential equations with indefinite characteristics
is often non-trivial and the general theory using  modern techniques
has only recently  been considered in
topologically trivial manifolds.
 \Ref{R Gardner, N Kamran,
{\it Hyperbolic Equations in the plane}.
},
 \Ref{D H Hartley, E D Fackerell R W Tucker,
{\it An Obstruction to the Integrability of a Class of Non-Linear
Wave Equations by 1-Stable Cartan Characteristics},
J Diff Equations, To Appear.
},
 \Ref { D H Hartley, P Tuckey, R W Tucker,
{\it Equivalence of Darboux and Gardner methods for Integrating hyperbolic
equations in the Plane},
Duke Math. J. To Appear
}.
Such techniques are relevant
for the general study of
(non-linear) equations that can arise on manifolds with a degenerate
geometry but need to be supplemented by further data to provide well posed
or interesting problems.

Kundt first
{\Ref {W Kundt {} Comm. Math. Phys. {\bf 4}
 (1967) 143 }}
discussed the non-existence of certain topologically non-trivial
spacetimes assuming that every geodesic is complete. Geroch
\Ref{R Geroch, J. Math. Phys. {\bf 11} (1970) 437} exploited the
notion of global hyperbolicity to reach a similar conclusion.

In this paper we consider two dimensional
 manifolds
with smooth (degenerate) metrics.
For the applications that we have in mind  we
require the existence of  asymptotically flat Lorentzian domains
 foliated by compact space-like hypersurfaces.

A non-trivial  two dimensional example is the trouser space. It may be
realized as a pair of trousers embedded in a Minkowskian spacetime of 3
dimensions  such that spacelike circles, disconnected at some time,
 become connected at another. Such a manifold cannot sustain a global
metric with a Lorentzian signature.
 The domain where the metric becomes  degenerate  depends on the
embedding but cannot be eliminated.

One of the fundamental issues that arises in describing fields on
such  manifolds is the dependence of the field equation on
the regularity of the metric tensor field. We adopt a pragmatic approach in
this paper and impose natural
conditions that enable us to construct non-singular
scalar fields that are globally $C^1$ in the presence of a degenerate
$C^\infty$ metric field.

An example of a two-dimensional
trouser-type manifold with a metric that is singular
at a single point may be found in
\Ref{H Ishikawa, Prog. Theor. Phys. {\bf
57} (1977) 339}.
In our approach we consider manifolds endowed
with a
smooth {\it everywhere regular} covariant metric tensor that is however
degenerate.
Such manifolds are therefore not {\it causally connected} \Ref{H Ishikawa, J.
Math. Phys.
Phys. {\bf 18} (1977) 2375},
\Ref {F J Tipler Ann. Phys. {\bf 108} (1977) 1  },
\Ref {S Hawking, R K Sachs, Comm. Math. Phys. {\bf 35} (1974) 287 }.

Since the trousers
 embedding is only for ease of visualisation one may equivalently
consider two (or more) cylinders surgically attached to a punctured 2-sphere.
It is then possible with the aid of smooth bump functions to endow such a
topological space with a metric that has Euclidean signature on the
punctured sphere and has  Lorentzian signature on a domain of the cylinders.
The transition between Euclidean and Lorentzian signature
is handled smoothly by the bump functions.

We describe below   regular global
 solutions to the massless scalar field
equation
$$d\star d\psi =0 \Eqno\label{\waveqn }$$
 on some  interesting spaces modeled on the above.
Here $d$ denotes the exterior derivative, $\star$ the Hodge map and $\psi$ a
complex scalar field.
Since
the punctured sphere is conformally flat one may use stereographic
projections as charts to map
 (the real part of) suitable complex analytic functions  that solve
 (1) on the punctured 2-plane,
to the punctured sphere. Since one can also solve (1) on the Lorentzian
cylinders it is possible to match them across the degeneracy curves to
construct a global solution.


\Section
{\bf  Differential Structure}

We first establish an atlas to describe a 2-dimensional
manifold, an example of which
 may be visualised as the embedding sketched in {\it Figure 1}.
We shall then endow this manifold with a $C^\infty$ metric that has
Euclidean signature where the cylinders are connected to the sphere and
Lorentzian signature elsewhere.

The manifold $M$  is constructed by first removing
     $n$ non-intersecting caps
$\{\Cap^\al\}_{\alpha = 1 \ldots n}$
from a topological sphere $S^2$.
Let $\S=S^2-\Union_\al\Cap^\al$.
Then from a set of $n$ cylinders, $\{\cylinder^\al\}_{\alpha = 1 \ldots n}$,
 smoothly attach
 each cylinder onto each of the  holes  made by removing  the caps
successively.


For each $\lal$  construct the  coordinate chart
$(\Ual , \Phi^\al)$ where

$$
\eqalign{
\Ual &= M - \{ {\rm all\ the\ cylinders\ except\ cylinder\ } \lal \} \cr
&=\S\union\cylinder^\al
}
$$
and
$$
\Phi^\al\colon \Ual \to \Real^2 \ : x \mapsto (\tau,\phi)
$$
where  $0\le\phi<2\pi$ and $-\infty<\tau\le\pi$ ,
( {\it see Figure 2 }).
For $(\tau,\phi)\in\Phil(\S)$ we adopt the standard $S^2$ metric for a
sphere of radius $R_1$:
$$
g^\al|_\S = \Rsq1\left(d\tau\otimes d\tau +
  \sin^2(\tau) d\phi\otimes d\phi\right)\,.
\Eqno
$$

For each cap let  $\tauS$ be the
angle subtended between the edge of the cap, the center of the sphere, and
the center of the cap $\cenal\in S^2$. For $\tau>\tauS$,
$(\tau,\phi)\in\Phil(\S)$
are standard
spherical coordinates.
Thus $\Phil(\Ual)$ has holes in it corresponding to the other
cylinders, but all the holes are in the region $\Phil(\S)$ .
When it is necessary to distinguish coordinates belonging to charts adapted
to different cylinders we shall append a chart label as a superscript to
the corresponding coordinates.
A  change of
coordinates from $(\tau^\al,\phi^\be)\in\Phil(\S)$ to
$(\tau^\be,\phi^\be)\in\Phi^\beta(\S)$
corresponds to an $SO(3)$
isometry of this metric. (This will be exploited with the aid
of Mobius transformations below).

For each $\lal$, choose a value of $\tauM$ such that
$\tauM<\tauS $.
Define the region $\Mal$
as follows:
$$
\Phil(\Mal) = \{(\tau,\phi):-\infty<\tau\le\tauM , 0\le\phi<2\pi\}
\Eqno
$$
On region $\Mal$  we adopt the standard flat Lorentzian
 metric for a cylinder of
radius $R_2$ given by
$$g^\al|_{\Mal} = - d\tau\otimes d\tau + \Rsq2 d\phi\otimes d\phi
\Eqno $$
Although $\dbydtau$ is directed towards the Euclidean region for each $\alpha$
we are free to choose an independent time-orientation for each
Lorentzian domain. With this in mind we are free to label cylinders as either
incoming or outgoing.
We now  smoothly connect the metric in regions ${\bf S}$ and $\Mal$
 with the aid of  bump functions.
Consider the function
$\bp\colon\Real \to \Real$ with $\bp \in C^\infty$ and
$\bp(x) = 0 , x \le 0$ and  $\bp(x) = 1 , x \ge 1$ with $\bp$ increasing.
There are standard techniques for constructing such a function
\Ref{ Th Br\"ocker, K J\"anich, Introduction to Differential Topology,  CUP
(1982)}.
Introduce
$$
f(\tau) = (\Rsq1 + 1) \bp \left( { \tau - \tauM \over \tauS - \tauM }
\right) -1
\Eqno $$
so that
$ f(\tau) = -1 $ for $ \tau<\tauM $ and
$ f(\tau) = \Rsq1 $ for $  \tau>\tauS $.

Define $\tauSig$ with
$\tauM<\tauSig<\tauS$
so that $f(\tauSig)=0$ and hence
$ f(\tau) < 0 $ for $ \tau<\tauSig $ and
$ f(\tau) > 1 $ for $  \tau>\tauSig $.

Further define
$$ h(\tau) = (\Rsq1 \sin^2(\tau) - \Rsq2)
\bp \left( { \tau - \tauM \over \tauS - \tauM }
\right)
+\Rsq2
\Eqno
$$
so that in  the region
$\Ual$ the metric is
given by:
$$ g^\al = f(\tau) d\tau\otimes d\tau + h(\tau)d\phi\otimes d\phi . \Eqno $$
Such a metric smoothly interpolates between  Lorentzian and Euclidean
regions.
Typical metric components are sketched in {\it Figure 3}.

We have adopted a metric with an  axial Killing symmetry in order to
expedite our discussion of (angular) momentum conservation below.

It is convenient to
introduce the regions $\Eal$ , $\Lal$ and the {\it rings}  ${\Sigma^\al}$~
defined in the table below.

\vskip 1em
\centerline{
\vbox{
\offinterlineskip
\hrule
\halign{
  &\vrule height 1.5em\hfil#\hfil
  &\vrule\hfil#\hfil
  &\vrule\hfil#\hfil
  &\vrule\hfil#\hfil
  &\vrule\hfil#\hfil
  &\vrule\hfil#\hfil
  \vrule
  \cr
&& $ \Phil(\Mal) $
&  $ \Phil(\Lal) $
&  $ \Phil(\Sigma^\al)$
&  $ \Phil(\Eal) $
&  $ \Phil(\S) $
&\cr
& $ \tau$
& $ -\infty<\tau\le\tauM $
& $ \tauM<\tau<\tauSig $
& $ \tau=\tauSig $
& $ \tauSig<\tau<\tauS $
& $ \tauS\le\tau<\pi $
&\cr
& $ \phi$
& $ 0\le\phi<2\pi $
& $ 0\le\phi<2\pi $
& $ 0\le\phi<2\pi $
& $ 0\le\phi<2\pi $
& $ 0\le\phi<2\pi $
&\cr
& $ f(\tau) $
& $ f(\tau)=-1 $
& $ f(\tau)<0 $
& $ f(\tau)=0 $
& $ f(\tau)>0 $
& $ f(\tau)=\Rsq1 $
&\cr
& $ h(\tau) $
& $ h(\tau)=\Rsq2 $
& $ h(\tau)>0 $
& $ h(\tau)>0 $
& $ h(\tau)>0 $
& $ h(\tau)=\Rsq1\sin^2(\tau) $
&\cr
& $ g $
& Flat Lorentzian
& Lorentzian
& Degenerate
& Euclidean
& Euclidean - Spherical
&\cr
 \omit\vrule height4pt&\omit\vrule height4pt&\omit\vrule height4pt
&\omit\vrule height4pt&\omit\vrule height4pt&\omit\vrule height4pt
&\omit\vrule height4pt&\cr
}
\hrule
}
}
so $\Ual={\bf S}\union\Eal\union\Lal\union\Mal$.
{Note:} The constants $\tauS,\tauSig,\tauM,R_2$ and the functions
$f(\tau),h(\tau)$ may be different for each cylinder $\alpha$ and will be
written
$\tauS^\al,\tauSig^\al,\tauM^\al,R^\al_2,f^\al(\tau),h^\al(\tau)$
respectively, when we wish to distinguish them.

It is worth noting that metrics of this type can readily be found that give
rise to bounded curvature scalars where the signature changes. For example
if
$f(\tauSig+t)=a_1 t+a_3 t^3 + \cdots$ and
$h(\tauSig+t)=b_0 + b_3 t^3 + b_4 t^4 + \cdots$ in the vicinity  of
$\tau=\tauSig+t$
then the magnitude of the curvature scalar
is $\displaystyle{9b_3 \over 2a_1b_0}$
at $\tau=\tauSig$,
where the metric becomes degenerate.

\Section{\bf   Matching Conditions}

We wish to construct functions $\psi\colon M\mapsto \Cmpx$
from  solutions to the equation \waveqn. Since the Hodge map is singular
where the metric tensor is degenerate we  restrict to solutions that
are $C^1$  across the rings $\Sigma^\alpha$.
By deriving {\waveqn } as  a local extremum of the
action
$$
\Lambda = \int_M d\psi\wedge\star\cnj{d\psi}
\Eqno\label\lagragian
$$
one recognises a hyperbolic wave equation in regions
$\bigcup_\al\Mal\union\Lal$ with Lorentzian signature and  an elliptic
(Laplace) Equation in the Euclidean region $\bigcup_\al{\bf S}\union\Mal$.
The closure of these regions intersect on the 1 dimensional rings $\Sigal$.
We assume that solutions to these local equations are continuous at these
rings:
$$ [\psi]_{\Sigal} = 0 \qquad \forall\lal
\Eqno\label\bddcona
$$
where, for any ring $\Sigma$, $[\omega]_\Sigma$ represents the discontinuity
$$
[\omega]_\Sigma =
\lim_{x\mapsto x_0\in\Sigma ,\, x\in\Eal} \omega_x \,\, -
\lim_{x\mapsto x_0\in\Sigma , \,x\in\Lal} \omega_x  .
\Eqno $$\label{\disc }
By demanding that the  contributions to the variations of the action cancel
on  $\Sigma^\al$  one derives  \Ref{T D Dray, C A Manogue, R W
Tucker, Phys. Rev. {\bf D48} (1993) 2587}\label\dray\
the natural junction conditions
$$
[{\Sigma^\al}^\star \star(d\psi)]_{\Sigma^\al} = 0\,.
\Eqno\label\bddconb
$$
Such conditions also arises naturally in a distributional
description \dray\ .
It is the purpose of this paper to find $C^1$  regular
 solutions on $M$ that satisfy
\waveqn, \bddcona\ and \bddconb.

\Section{\bf  Mapping the Euclidean Domain into the Complex Plane}

Since the manifold is two dimensional and \waveqn\ is conformally covariant
under scalings of the Euclidean metric it is natural to use the complex
plane. However although one may map most of a sphere to the complex plane
using the canonical stereographic mapping, we note that the entire Euclidean
region  $\ME$ of our problem comprises the compact set
$$
\ME=\S\union\left(\Union_\al \Eal\union\Sigal \right)\subset M \,.
$$
It is useful therefore to  first consider an
injection of the Euclidean region into the sphere
$$
\Proj:\ME\mapsto S^2
$$
 given by
$$
\Proj|_{\S} : \S\mapsto S^2
\hbox{ is the identity}
\Eqno\label\defprojS
$$
and
$$
\eqalign{
&\Proj|_{\Eal} : \Eal \mapsto S^2 \cr
&\left(\theta=2 \arctan \left( {\rho e^{\Gt} \over 2 R_1 } \right) ,
\phi = \phi \right)
}
\Eqno\label\defprojE
$$
where $(\theta,\phi)$ are the usual spherical coordinates for $S^2$ about the
point $s_\al$ labeling the centre of cylinder $\alpha$.

The pair  $(\tau,\phi)$ denote the coordinates of the chart
$(\Ual,\Phil)$ restricted to the Euclidean region and
$$
\Gt = \int_{\tauSig}^\tau \left|
{f\left(\tau'\right) \over h\left(\tau'\right)} \right|^{1 \over 2}
d\tau' \,,
\qquad
\rho = 2 R_1 \tan\!\left({\textstyle {\tauS \over 2}}\right)
e^{-G(\tauS)}
\Eqno\label\defrho
$$
The constant $\rho$ and the function
$G(\tau)$ may be different for each cylinder $\alpha$ and will be
written
$\rho_\al$,  $G^\al(\tau)$
respectively when we wish to distinguish them.

The definition \defprojE\ of $\Proj|_{\Eal}$ extends naturally to
$\Proj|_{\Eal\union\S}$ and agrees with $\Proj|_\S$ of definition
\defprojS\ :
for $(\tau,\phi)\in\Phil(\S)$
$$
\eqalign{
2 R_1 \tan(\textstyle{\theta \over 2} )
&=
\rho e^{\Gt} \cr
&= 2 R_1 \tan(\textstyle{ \tauS \over 2})
e^{\Gt-G(\tauS) }\cr
&=
2 R_1 \tan({\textstyle{ \tauS \over 2}})
\exp\left(\int^\tau_{\tauS} {1 \over \sin(\tau')} d\tau \right) \cr
&=
2 R_1  \tan(\textstyle{ \tauS \over 2})
\exp\left(\log(\tan(\textstyle{ \tau \over 2}))-
          \log(\tan(\textstyle{ \tauS \over 2}) )\right) \cr
&=
2 R_1 \tan(\textstyle{ \tau \over 2}) . \cr
\,}
$$

We may now map the entire Euclidean domain $\ME$ into
$\Cminf=\Cmpx\union\{\infty\}$, the one point compactified
complex plane with an $S^2$ topology.
Recall that for each $\lal$, $\cenal\in S^2$ is the
center of
the $\Cap^\al$ (removed in the construction of the manifold $M$).
Denote by  $\oppal\in S^2$  the point
antipodal to $\cenal$.
For each cylinder $\lal$ there exists a stereographic projection
\hbox{$\prohal:S^2-\{\oppal\}\mapsto\Cmpx$}
 such that $\prohal(\cenal)=0\in\Cmpx$~.
We  extend this to
$$
\eqalign{
\prohal&:S^2\mapsto\Cminf \cr
&:(\theta,\phi)\mapsto
2 R_1 e^{i\phi} \tan(\textstyle{ \theta \over 2}) \cr
&:\oppal\mapsto\infty
}
$$
We  now combine these maps
$$
\eqalign{
&\projal:\ME\mapsto\Cminf \cr
&\projal = \prohal\circ\Proj
}
$$
to construct the required projection of $\ME$ into $\Cminf$.
If $x\in\S\union\Eal=\ME\inter\Ual$ then in the $(\tau,\phi)$ chart we
have $\Phil(x)=(\tau,\phi)$ with
$\projal(x)=\rho e^{\Gt+i\phi}.$
Examples of the maps $\projal$, $\prohal$ and $\Proj$ are sketched in
{\it Figure 4}.

Since $\projbe\projal\tmone:\projbe(\ME)\mapsto\projal(\ME)$ is the same as
 $\prohbe\prohal\tmone:\Cminf\mapsto\Cminf$ where both are defined we have
 a mapping between two different projections of a sphere. Such a
mapping may be represented by a special
Mobius Transformation
representing  an $SO(3)$ rotation of the sphere:
$$
\prohbe\prohal\tmone(z)
=
e^{i a_3^\albe}
\left(
{ -2R_1\tan(\half a_2^\albe) + z e^{i a_1^\albe}
\over\displaystyle
1 + {\strut \tan(\half a_2^\albe)  z e^{i a_1^\albe} \over 2R_1 }
}
\right)
\Eqno\label\projdef
$$
where $\{a_1^\albe,a_2^\albe,a_3^\albe\}\in\Real^3$~,
$$
2 R_1\tan(\half a_2^\albe) e^{-i a_1^\albe} = \prohal(\cenbe)
\qquad{\rm and}\qquad
-2 R_1\tan(\half a_2^\albe) e^{ i a_3^\albe}
= \prohbe(\cenal) \,.
$$

The parameters $\{a_1^\albe,a_2^\albe,a_3^\albe\}$ represent the Euler
angles  of
the $SO(3)$ rotation.
 In terms of the coordinates on the sphere, the pair
$(a_2^\albe,-a_1^\albe)$ denote the $(\theta,\phi)$ coordinates of the
point $\cenbe$ with respect to the spherical coordinate system about
$\cenal$~. Similarly
$(-a_2^\albe,a_3^\albe)$  are the $(\theta,\phi)$ coordinates of $\cenal$
with respect
to the spherical
coordinate system about $\cenbe$.
 By looking at the inverse of $\prohal\prohbe\tmone$ we note
that $a_1^{\albe}=-a_3^{\beal}$~, $a_2^{\albe}=a_2^{\beal}$~, and
$a_3^{\albe}=-a_1^{\beal}$~.
Since we are free to choose the origin of the $\phi$
coordinate for each $\lal$ we may use this freedom to make some of the
$a_1^{\albe},a_3^{\albe}$
to vanish. ($a_2^{\albe}$ is fixed by the location of holes on the manifold.)
For example if there are just two cylinders we may choose $a_1^{(12)}=0$ and
$a_3^{(12)}=0$ by
the choice of the origins of $\phi^{(1)}$ and $\phi^{(2)}$.
For a third cylinder, we may take $a_3^{(13)}=0$ by the choice of
origin of $\phi^{(3)}$,
but $a_1^{(23)},a_3^{(23)},a_1^{(13)}$ must now be calculated using
$\prohn2\prohn3\tmone = (\prohn3\prohn1\tmone)\circ(\prohn1\prohn2\tmone)$
and in general this will be non zero.

\Section{\bf  Construction of Global Solutions }

{\bf Theorem (1.1)}
\newline
Given the analytic functions
$\psi^\al_\pm:\projal(\ME)\mapsto\Cmpx$
and a set of constants $A^\del,B^\del\in\Cmpx$ for each
cylinder~$\delta$ satisfying $\sum_\delta A^\del = 0$~,
then
$\psi|_\ME:\ME\mapsto\Cmpx$ is a solution to \waveqn\
 given by
$$
\eqalign{
\psi|_\ME(x) &=
\psi^\al_+(\projal(x)) + \cnj{\psi^\al_-(\projal(x))}
+ \sum_{\cendel\ne\oppal} A^\del \log|\projal(x)-\prohal(\cendel)|
+ B^\al
\,.}
\Eqno\label\solE
$$
The sum $\sum_{\cendel\ne\oppal} $
here is over all caps $\delta$ excluding the cap whose centre is
$\oppal$, should it exist. (This is because if $\cendel=\oppal$ then
$\log|\projal(x)-\prohal(\cendel)|=\log|\projal(x)-\prohal(\oppal)|
=\log(\infty).$~)

{\bf Theorem (1.2)}
\newline
Given periodic functions $\psi_\pm^{L\al}\colon S^1 \mapsto \Cmpx$
then in the Lorentzian region  of the cylinder $\lal$,
a solution $\psi|_\LuM:\LuM\mapsto\Cmpx$ is given in the chart
$\Phil(x)=(\tau,\phi)$ by
$$
\psi|_\LuM(x) =
\psi_+^{L\al}\left(\phi+G^\al(\tau)\right) +
\psi_-^{L\al}\left(\phi-G^\al(\tau)\right)
-
A^\al G^\al(\tau)
+
A^\al\log(\rho_\al)+B^\al \,.
\Eqno\label\solL
$$

{\bf Theorem (1.3)}
\newline
If $\psi_\pm^\al,\psi_\pm^{L\al}$ satisfy the conditions
$$
\eqalign{
\psi_+^{L\al}(\phi) &= \half(1+i)(\hh\psi_+^\al(\rho e^{i\phi}) -
   i\cnj{\hh\psi_-^\al(\rho e^{i\phi})}) \cr
\psi_-^{L\al}(\phi) &= \half(1-i)(\hh\psi_+^\al(\rho e^{i\phi}) +
   i\cnj{\hh\psi_-^\al(\rho e^{i\phi})}) \cr
}
\Eqno\label\solconver
$$
where
$$
\eqalign{
\hh\psi_\pm^\al &: \projal(\Eal\union\Sigal)\mapsto\Cmpx \cr
\hh\psi_\pm^\al(z) &=
\psi_\pm(z)+
\sum_{\cendel\ne\cenal,\oppal} A^\del \log|z-\prohal(\cendel)|
\qquad\forall z\in\projal(\Eal\union\Sigal)
}
\Eqno\label\defhhpsi
$$
then $\psi|_{\Ual}$ satisfies the
conditions \bddcona\ and \bddconb.
The sum $\sum_{\cendel\ne\cenal,\oppal}$ here is over all caps $\delta$
excluding the cap $\alpha$ but also excluding the cap whose centrex is
$\oppal$, should it exist.

{\bf Theorem (1.4)}
\newline
Any solution to  \waveqn, \bddcona\ and \bddconb\ can be written
in this form.

{\bf Theorem (1.5)}
\newline
Under change of coordinates $\Phil$ to $\Phibe$
$$
\eqalign{
\psi^\be_+ \circ \projbe &= \psi^\al_+ \circ \projal \cr
\psi^\be_- \circ \projbe &= \psi^\al_- \circ \projal \cr
}
\Eqno\label\conME
$$
$$
B^\be = B^\al +
\sum_{\cendel\ne\oppal,\oppbe}\!\!\! A^\del
\log|\prohal(\cendel)-\prohal(\oppbe)|
+ A^\beinf \log |\prohal(\oppbe)| +
A^\beinf \log |\projbe(\cenal)-\prohbe(\oppal)|\,.
$$

The sum  $\sum_{\cendel\ne\oppal,\oppbe}$ here is over all caps $\delta$
excluding the caps  whose centres are either  $\oppal$  or $\oppbe$ should
they exist. In the case when $\oppbe$ is the centrex of a cap (say $\oppbe$
is $\cendel$) let $A^{\beinf}$ be the corresponding constant $A^\delta$.


{\bf Proof of Theorem (1.1) }
\newline
If $\oppal\in\projal(\ME)$ then since $\sum_\del A^\del=0$,
$\psi|_\ME$ is a well defined function. Also since $\ME$ is compact, so is
$\projal(\ME)$, hence $\psi_\pm(\projal(\ME))$ is closed and bounded. Thus
$\psi:M\mapsto\Cmpx$ is non singular.
In the Euclidean region let
$\Phil(x)=(\tau,\phi)$ where $x\in\ME\inter\Ual=\S\Union\Eal$. Then
$\projal(x)=\rho e^{i\phi+\Gt}$.
It is straightforward to verify that this satisfies \waveqn\ in
${\ME\inter\Ual}$. \qed

{\bf Proof of Theorem (1.2) }
\newline
Trivial. \qed

{\bf Proof of Theorem (1.3) }
\newline
For this part of the proof we  drop the $\lal$ and write
$\psi_\pm^\al=\psi_\pm$ , $\psi_\pm^{L\al}=\psi_\pm^L$ , $\rho_\al=\rho$,
 $G^\al(\tau)=\Gt$~.
In the Euclidean region  let
$\Phil(x)=(\tau,\phi)$ where $x\in\ME\inter\Ual=\S\Union\Eal$. Then
$$
\psi|_{\ME\inter\Ual}(x) =
\psi_+(\rho e^{i\phi+\Gt}) + \cnj{\psi_-(\rho e^{i\phi+\Gt})}
+ \sum_\del A^\del \log|\rho e^{i\phi+\Gt}-\prohal(\cendel)|
+ B^\al
\,.$$
Putting $\tau=\tauSig$, since $G(\tauSig)=0$,  the continuity
condition \bddcona\ is satisfied.

We next pull back this Euclidean solution to $\Sigal$.
Since $\projal(\Eal\union\Sigal)$ is an
annulus about $0\in\Cmpx$,
\hbox{$\log(z-\prohal(\cendel))$}
is a well defined function for
all $\del\ne\lal$. Thus $\hh\psi_\pm$ is well defined and
$$
\psi|_{\Eal}(x)=
\hh\psi_+(\rho e^{i\phi+\Gt}) + \cnj{\hh\psi_-(\rho e^{i\phi+\Gt})}
+A^\al\Gt +A^\al\log(\rho) + B^\al
\,. $$
The Hodge dulls are
$$
\star d\tau = {1 \over G'(\tau)} d\phi
{\rm\ \ and\ \ }
\star d\phi = -G'(\tau) d\tau
$$
hence
$$
\star d\hh\psi_\pm(\rho e^{i\phi+\Gt})
=
\hh\psi_\pm {}'
(\rho e^{i\phi+\Gt})
\rho e^{i\phi+\Gt}
(-iG'(\tau) d\tau +  d\phi)
\Eqno\label\hhpsiE
$$
$$
\Sigma^\star \star d\hh\psi|_\EuS
=
\left(
\rho e^{i\phi} \hh\psi_+'(\rho e^{i\phi})
+
\rho e^{-i\phi} \cnj{\hh\psi_-'(\rho e^{i\phi})}
+
A^\al
\right) d\phi
\,.$$

Next consider the solution in  the Lorentzian region $x\in\Lal\union\Mal$.
The Hodge duals here are
$$
\star d\tau = -{1 \over G'(\tau)} d\phi
{\rm\ \ and\ \ }
\star d\phi = -G'(\tau) d\tau
$$
hence
$$
\eqalign{
\star d\psi|_\LuM
&=
\star d\psi_+^L(\phi+\Gt) +
\star d\psi_-^L(\phi-\Gt)
- A^\al\star d\Gt
\cr
&=
\half(1+i)\star d\left(
\hh\psi_+\left(\rho e^{i(\phi+\Gt)}\right) -
i\cnj{\hh\psi_-\left(\rho e^{i(\phi+\Gt)}\right)}
\right) \cr
& \quad +
\half(1-i)\star d\left(
\hh\psi_+\left(\rho e^{i(\phi-\Gt)}\right) +
i\cnj{\hh\psi_-\left(\rho e^{i(\phi-\Gt)}\right)}
\right)
+A^\al d\phi
\cr
&=
\half(1+i)\rho(-id\phi - iG'(\tau) d\tau)
e^{i(\phi+\Gt)}
\hh\psi_+'\left(\rho e^{i(\phi+\Gt)}\right)
\cr
&\quad -
i\half(1+i)\rho(id\phi + iG'(\tau) d\tau)
e^{-i(\phi+\Gt)}
\cnj{\hh\psi_+'\left(\rho e^{i(\phi+\Gt)}\right)}
\cr
&\quad +
\half(1-i)\rho(id\phi - iG'(\tau) d\tau)
e^{i(\phi-\Gt)}
\hh\psi_+'\left(\rho e^{i(\phi-\Gt)}\right)
\cr
& \quad +
i\half(1-i)\rho(-id\phi + iG'(\tau) d\tau)
e^{-i(\phi-\Gt)}
\cnj{\hh\psi_+'\left(\rho e^{i(\phi-\Gt)}\right)}
+A^\al d\phi
}
$$
\vskip 1em
$$
\eqalign{
\Sigma^\star \star d\psi|_\LuM
&=
\half(1+i)(-i)\;d\phi\;\rho
e^{i\phi}
\hh\psi_+'\left(\rho e^{i\phi}\right)
-i\half(1+i)i\;d\phi\;\rho
e^{-i\phi}
\cnj{\hh\psi_-'\left(\rho e^{i\phi}\right)}
\cr
&+
\half(1-i)i\;d\phi\;\rho
e^{i\phi}
\hh\psi_+'\left(\rho e^{i\phi}\right)
+i(1-i)(-i)\;d\phi\;\rho
e^{-i\phi}
\cnj{\hh\psi_-'\left(\rho e^{i\phi}\right)}
 +  A^\al \;d\phi\;
\,.}
$$
Hence
$$
\Sigma^\star \star d\psi|_\LuM =
\Sigma^\star \star d\psi|_\EuS
$$
so the condition on the first derivatives \bddconb\ is satisfied. \qed

In order to prove {\bf Theorem (1.4)} we need the following lemmas:

{\bf Lemma (1.4.1)}
\newline
Let $D\subset\Cminf$ be a closed pathwise connected subset and
$\psi\colon D\mapsto\Real$ a solution of  Laplace's equation $d\star d\psi=0$,
where $\Cmpx$ has the standard $\Real^2$ Euclidean metric.
Each hole in $D$ is a connected open subset $C^\al$,
so that
$\Union_\al C^\al=\Cminf-D$ is a disjoint union.
For each hole $C^\al$ choose
$a^\al\in C^\al$.
Then there exists an analytic function $f\colon D\mapsto\Cmpx$ and
constants $A^\al\in\Real$
such that we can write
$$
\psi(z)={\rm Re} (f(z)) + \sum_\al A^\al\log(|z-a^\al|)
\,.$$

{\bf Proof of Lemma (1.4.1)}
\newline
If $D$ were simply connected then writing $z = x+iy$ and
$f(x+iy)=\psi(x+iy)+i\varphi(x+iy)$ the Cauchy Riemann equations imply
$$
{\partial\psi\over\partial x} = {\partial\varphi\over\partial y}
{\rm\ \ \ and\ \ }
{\partial\psi\over\partial y} = -{\partial\varphi\over\partial x}
\ .$$
Given $\psi$ we can use these to determine $\varphi$ up to a constant.

Since $D$ is not simply conected, $\varphi$ need not be single valued but
can at worst be a function of the winding numbers $n^\al\in {\bf Z}$ around
each hole
$C^\al$. Suppose $\varphi$ increase by $2\pi n^\al A^\al$ as it winds
round $C^\al$ once.  By subtracting $A^\al\log(z-a^\al)$ from $f(z)$ we are
left with a single valued function in a neighbourhood of the hole $\lal$.
Hence
$$
\widehat f(z) \mathrel{\mathop=^{\rm def}}
f(z) - \sum_\al A^\al\log(z-a^\al) =
\psi(z) + i\varphi(z) - \sum_\al A^\al\log(z-a^\al)
$$
is now single valued.
Taking the real part we prove the lemma.
\qed

{\bf Lemma (1.4.2)}
$$\sum_\del A^\del=0\,.$$

{\bf Proof of Lemma (1.4.2)}
\newline
If $\infty\not\in D\subset\Cminf$,
then $\infty\in\Cminf-D$ and $\infty\in C^\delta$ for one of the $\del$.  By
considering the variation of $\varphi$ along a contour $\Gamma$ just inside
$\Cmpx-C^\delta$, so that $\Gamma$ encloses all the other holes
$\{C^\al\}_{\al\ne\del}$ we see that $\sum_{\al\ne\del}A^\al=-A^\del$.

If $\infty\in D\subset\Cminf$,
then $\psi\colon D\mapsto\Real$ is bounded as $z\to\infty$ so again
considering a contour that $\Gamma$ encloses all the holes, we obtain the
same condition.
\qed

{\bf Lemma (1.4.3)}
\newline
If $\psi\colon \Real\times S^1\mapsto\Real$ satisfies the hyperbolic equation
$d\star d\psi=0$,
where $\Real\times S^1$ has the standard Lorentzian metric, then
there exist functions $\psi_\pm \colon S^1\mapsto\Real$ and a constant
$A\in\Real$ such that
$$
\psi(\tau,\phi) = \psi_+(\phi+\tau) +  \psi_-(\phi-\tau) + A\tau
$$
where $(\tau,\phi)$ are coordinates for $\Real\times S^1 .$

{\bf Proof of Lemma (1.4.3)}
\newline
Trivial. \qed

With these lemmas we return to the proof of {\bf Theorem (1.4).}

Suppose  $\psi: M \mapsto \Cmpx$ satisfies equation
\waveqn\ .
Since $\projal$  is a conformal mapping
$\psi\circ\projal{}^{-1}:\projal(\ME)\mapsto\Cmpx$ satisfies Laplace's
 equation.
Then
${\Re}(\psi\circ\projal{}^{-1}),
{\Im}(\psi\circ\projal{}^{-1}):\projal(\ME)\mapsto\Real$ each
satisfy Laplace's equation.
Also $\projal(\ME)$ is a pathwise
connected closed subset of $\Cminf$.
First assume that $\prohal(\cendel)\ne\infty$ for all $\cendel$.
Choose $a^\del=\prohal(\cendel)$.
Therefore from {\bf lemma (1.4.1)}
there exist  analytic functions
$f_1,f_2\colon\projal(\ME)\mapsto\Cmpx$ and
constants $A_2^\al,A_1^\al\in\Real$
such that we can write (with $x\in\ME$ and $z=\projal(x)$~)~:
$$
\eqalign{
{\Re}(\psi(x))={\rm Re} (f_1(z)) +
              \sum_{\cendel} A_1^\del\log(|z-\prohal(\cendel)|)
\cr
{\Im}(\psi(x))={\rm Re} (f_2(z)) +
              \sum_{\cendel} A_2^\del\log(|z-\prohal(\cendel)|)
\,.}
$$
Let
$$
\eqalign{
\psi^\al_+(z) = f_1(z) + i f_2(z) \cr
\psi^\al_-(z) = f_1(z) - i f_2(z) \cr
A^\del = A_1^\del +  A_2^\del \,.
}
$$
Then
$$
\psi|_\ME(x) =
\psi^\al_+(\projal(x)) + \cnj{\psi^\al_-(\projal(x))}
+ \sum_\cendel A^\del \log|\projal(x)-\prohal(\cendel)|
+ B^\al \,.
\Eqno\label\solEnoopp
$$
If on the other hand $\oppal=\cenbe$ for some $\be$ then
we are left with an extra term $A^\alinf\log|z-a^\be|$ where
$z=\prohal(x)$.
However $\log|z-a^\be|=\log|1-a^\be/z|+\log|z|$ and $\log(1-a^\be/z)$ is
analytic on $\projal(\ME)$.
So \solEnoopp\ becomes
$$
\eqalign{
\psi|_\ME(x) = &
\left(
\psi^\al_+(\projal(x)) + A^\alinf\log(\projal(x))
\right) +
\left(
\cnj{\psi^\al_-(\projal(x)) + A^\alinf\log(\projal(x))}
\right)
\cr + & \qquad
\sum_{\cendel\ne\cenal,\oppal}
A^\del \log|\projal(x)-\prohal(\cendel)|
+ (A^\al+A^\alinf)\log|\prohal(x)|
+ B^\al
}
$$
and {\bf Theorem (1.4)} is true in the Euclidean region.
Furthermore  the Lorentzian region on one cylinder $\Lal\union\Mal$ is
conformal
 to the  cylinder $\Real\times S^1$, so  proving {\bf Theorem (1.4)} for
such a  region is equivalent to {\bf Lemma (1.4.3)}

Finally we wish to show that the joining conditions
\bddcona\  and
\bddconb, imply the relationship \solconver.
Let $A^L,B^L$ be the $A^\al,B^\al$ of equation \solL\ and $A^E,A^E$ be the
$A^\al,B^\al$ of equation \solE.
If $\tau=\tauSig$ then
\bddcona\ implies
$$
\psi_+^L(\phi) + \psi_-^L(\phi) + B^L=
\psi_+(\rho e^{i\phi}) + \cnj{\psi_+(\rho e^{i\phi})} + B^E \,.
\Eqno\label\eqntaoO
$$
Also from \bddconb\ we have
$$
\left(\psi_+^L{}'(\phi) + \psi_-^L{}'(\phi) + A^L
\right) d\phi=
\left(
\rho e^{i\phi}\psi_+{}'(\rho e^{i\phi}) +
\cnj{\rho e^{i\phi}\psi_+{}'(\rho e^{i\phi})} + A^E
\right) d\phi \,.
$$
Integrating this gives
$$
\psi_+^L(\phi) + \psi_-^L(\phi) + A^L\phi
=
-i\psi_+(\rho e^{i\phi})
+i\,\cnj{\psi_+(\rho e^{i\phi})} + A^E\phi + C
$$
and solving these equations we get \solconver\ up to the choice of
$B$.  \qed


{\bf Proof of Theorem (1.5)}
\newline
By substituting $z'=\projbe\projal{}^{-1}(z)$ where $z=\projal(x)$ in
\solE\ , the first two of \conME\ are automatically satified.

If $\Mob:\Cminf\mapsto\Cminf$ is a Mobius transformation then for $a,b\in\Cmpx$
$$
{ a - b \over a - \Mob(\infty) }
=
{ \Mob^{-1}(b) -  \Mob^{-1}(a) \over \Mob^{-1}(b) -  \Mob^{-1}(\infty) }\,.
\Eqno
$$
Substituting $\Mob=\projal\projbe{}^{-1}$ , $a=\projal(y)$ , $b=\projal(x)$
then $\Mob(\infty)=\prohal(\oppbe)$ and $\Mob^{-1}(\infty)=\prohbe(\oppal)$
in this gives
$$
{ \projal(y)-\projal(x) \over \projal(y) - \prohal(\oppbe) }
=
{ \projbe(x)-\projbe(y) \over \projbe(x) - \prohbe(\oppal) }
\,.$$
Writing  $y=\cendel$ and taking logs
 gives the relation:
$$
\log|\projal(x)-\prohal(\cendel)|
=
\log|\projbe(x)-\prohbe(\cendel)|
- \log|\projbe(x)-\prohbe(\oppal)|
+ \log|\prohal(\cendel)-\prohal(\oppbe)| \,.
$$
This is
valid except for the case when $\cendel\ne\oppal$ and $\cendel\ne\oppbe$
$\forall\delta$ since then $\prohal(\oppal)=\infty$.
For this case we first  note that
$$
(\Mob(\infty)-b)(\Mob^{-1}(b)-\Mob^{-1}(\infty))=K
\Eqno$$\label\keqn
where $K$ is independent of $b$. Subtracting \keqn\ with $b=\projal(x)$
from \keqn\ with  $b=0$  gives
$$
(\prohal(\oppbe)-\projal(x))(\projbe(x)-\prohbe(\oppal))
=
\prohal(\oppbe)(\projbe(\cenal)-\prohbe(\oppal))
$$
and hence
$$
\log |\prohal(\oppbe)-\projal(x)| =
- \log |\projbe(x)-\prohbe(\oppal)| +
\log |\prohal(\oppbe)| +
\log |\projbe(\cenal)-\prohbe(\oppal)|
\,.$$
Thus
$$
\eqalign{
&\sum_{\cendel\ne\oppal} A^\del \log|\projal(x)-\prohal(\cendel)|
\cr &=
\sum_{\cendel\ne\oppal,\oppbe}\!\!\! A^\del \log|\projal(x)-\prohal(\cendel)|
+
A^\beinf \log|\projal(x)-\prohal(\oppbe)|
\cr &=
\sum_{\cendel\ne\oppal,\oppbe}\!\!\!A^\del
\log|\projbe(x)-\prohbe(\cendel)|
-\sum_{\cendel\ne\oppal,\oppbe}\!\!\!A^\del
\log|\projbe(x)-\prohbe(\oppal)|
\cr &\quad
+\sum_{\cendel\ne\oppal,\oppbe}\!\!\!A^\del
\log|\prohal(\cendel)-\prohal(\oppbe)|
-A^\beinf \log |\projbe(x)-\prohbe(\oppal)| +
A^\beinf \log |\prohal(\oppbe)| +
\cr &\quad
A^\beinf \log |\projbe(\cenal)-\prohbe(\oppal)|
\,.}$$
Now $\psi$ and $\projal(\ME)$ satify one or other of the conditions for
{\bf Lemma (1.4.2)} ($\sum_\del A^\del=0$) so
$$
-\sum_{\cendel\ne\oppal,\oppbe}\!\!\! A^\del \log|\projbe(x)-\prohbe(\oppal)|
=
(A^\alinf + A^\beinf)\log|\projbe(x)-\prohbe(\oppal)|
$$
$$
\eqalign{
&\sum_{\cendel\ne\oppal,\oppbe}\!\!\!A^\del
\log|\projbe(x)-\prohbe(\cendel)|
-\sum_{\cendel\ne\oppal,\oppbe}\!\!\!A^\del
\log|\projbe(x)-\prohbe(\oppal)|
\cr & \qquad
-A^\beinf \log |\projbe(x)-\prohbe(\oppal)|
=\!\!\!
\sum_{\cendel\ne\oppbe}A^\del
\log|\projbe(x)-\prohbe(\cendel)|
}
$$
$$
\eqalign{
\sum_{\cendel\ne\oppal} A^\del \log|\projal(x)-\prohal(\cendel)|
= &\!\!\!\!
\sum_{\cendel\ne\oppbe} A^\del \log|\projbe(x)-\prohbe(\cendel)|
+\!\!\!\!\!\!\sum_{\cendel\ne\oppal,\oppbe}\!\!\! A^\del
\log|\prohal(\cendel)-\prohal(\oppbe)|
\cr&
+ A^\beinf \log |\prohal(\oppbe)| +
A^\beinf \log |\projbe(\cenal)-\prohbe(\oppal)|
\,.}
$$
Finally
$$
\sum_{\cendel\ne\oppal} A^\del \log|\projal(x)-\prohal(\cendel)|
+ B^\al
=
\sum_{\cendel\ne\oppbe} A^\del \log|\projbe(x)-\prohbe(\cendel)|
+ B^\be
$$
and {\bf Theorem (1.3)} is satified. \qed

Observe that in the flat Minkowski regions $\Mal$,
with $\Phil(x)=(\tau,\phi)$, \solL\ becomes
$$
\psi|_\Mal(x) =
\psi_+^{L\al}\left(\phi+{\tau-\epsilon \over R_2}\right) +
\psi_-^{L\al}\left(\phi-{\tau-\epsilon \over R_2}\right)
-
A^\al {\tau-\epsilon \over R_2}
+
A^\al\log(\rho_\al)+B^\al
$$
where
$$
\epsilon=\epsilon^\al=
G^\al(\tauM^\al)-\tauM^\al \,.
\Eqno\label\defeps
$$
We also note that the global solution
is $C^1$ and piecewise  $C^\infty$.
\def\tt{\widetilde}

\Section{{Field Energy and Momentum}}

For a field configuration $\psi$ and a local vector field $X$ on $M$
we define the forms
$$
\Tau_X =
i_X d\psi\wedge\star d\psi +
d\psi\wedge i_X\star d\psi \,.
\Eqno
$$
In Lorentzian regions where X is one of the
 Killing vectors  $\dbydtau$ or $\dbydphi$ these identify energy and
(angular) momentum density 1-forms  {\Ref{I M Benn, R W Tucker, An
Introduction to Geometry and Spinors with Applications in Physics. Adam
Hilger (1988) }} .
Integrating these over a ring of constant $\tau$ in a Lorentzian region
gives an energy and
(angular)
momentum appropriate to that hypersurface.
$$
\eqalign{
\eng^\al(\tau) &=
\int_0^{2\pi}
\left(
(i_{\dbydtau} d\psi) \star d\psi +
d\psi (i_{\dbydtau}\star d\psi)
\right)\cr
\ang^\al(\tau) &=
\int_0^{2\pi}
\left(
(i_{\dbydphi} d\psi) \star d\psi +
d\psi (i_{\dbydphi}\star d\psi)
\right)
\,.}
\Eqno\label\defeng
$$
We may define similar quantities by
integrating  over a ring of constant $\tau$ (where it exists)
in the Euclidean region.
Where the metric has axial symmetry, i.e. on $\Eal \union \Sigal \union \Lal
\union \Mal $,
$$
\lie_{\dbydphi} g = 0
\Eqno
$$
where $\lie$ denotes the Lie derivative. It follows that
the momentum $\ang^\al(\tau)$ is  a constant in all regions where
the metric is nondegerate.
Furthermore $\dbydtau$ is also a Killing vector
in the  flat Lorentzian regions, $\Mal$, of the manifold:
$$
\lie_{\dbydtau} g = 0
\,.\Eqno
$$
In these regions we therefore have two constants of the motion
$\eng^{L\al}$ and
$\ang^{L\al} $ since \defeng\ are independent of $\tau$.
We now compute the forms \defeng\  corresponding to the Lorentzian
 solutions \solL\ :
$$
\eqalign{
\eng^{L\al}(\tau)
=
4\!\left(\! {-f(\tau)\over h(\tau)}\! \right)^{1 \over 2}\!\!
&\bigg(\!\!
\int_0^{2\pi}\!\!\!
\Big(\!
\psi_+^{L\al}{}'(\phi\!+\!\Gt)
\;\cnj{\psi_+^{L\al}{}'(\phi\!+\!\Gt) }
+
\cr &
\psi_-^{L\al}{}'(\phi\!-\!\Gt)
\;\cnj{\psi_-^{L\al}{}'(\phi\!-\!\Gt)}
\Big)
d\phi
+
\pi A^\al \cnj{A^\al}
\bigg)
}
\Eqno\label\engfunL
$$
and
$$
\eqalign{
\ang^{L\al} =
4
\int_0^{2\pi}
&\Big(
\psi_+^{L\al}{}'(\phi-\Gt)
\;\cnj{\psi_+^{L\al}{}(\phi-\Gt)}
-
\psi_-^{L\al}{}'(\phi+\Gt)
\;\cnj{\psi_-^{L\al}{}'(\phi+\Gt)}
\Big)
d\phi
}
\,.
\Eqno\label\angfunL
$$
Since each  integrand is periodic in $\phi$ these may be simplified  by the
substitutions  {$\phi\to\phi\pm\Gt$:}
$$
\eqalign{
\eng^{L\al}(\tau)
=
4\left( {-f(\tau)\over h(\tau)} \right)^{1 \over 2}
\bigg(
\int_0^{2\pi}
&\Big(
\psi_+^{L\al}{}'(\phi)
\;\cnj{\psi_+^{L\al}{}'(\phi)}
+
\psi_-^{L\al}{}'(\phi)
\;\cnj{\psi_-^{L\al}{}'(\phi)}
\Big)
d\phi
+
\pi A^\al \cnj{A^\al}
\bigg)
}
\Eqno\label\engLL
$$
$$
\eqalign{
\ang^{L\al} &=
4
\int_0^{2\pi}
\Big(
\psi_-^{L\al}{}'(\phi)
\;\cnj{\psi_-^{L\al}{}'(\phi)}
-
\psi_+^{L\al}{}'(\phi)
\;\cnj{\psi_+^{L\al}{}'(\phi)}
\Big)
d\phi \,.
}
\Eqno\label\angLL
$$

For the solution in the  Euclidean region $\Eal$:
$$
\eqalign{
\eng^{E\al}(\tau) &=
-4\!\left( {f(\tau)\over h(\tau)} \right)^{1 \over 2}\!\!\!
\Re\!\left(\! 2\rho^2\!\!
\int_0^{2\pi}
\hh\psi^\al_+{}'\!\left(\rho e^{i\phi+\Gt}\right) \;
\hh\psi^\al_-{}'\!\left(\rho e^{i\phi+\Gt}\right)
e^{2(i\phi+\Gt)}
d\phi
+ \pi A^\al \cnj{A^\al}
\right)
}\Eqno
$$
and
$$
\ang^{E\al} =
8\rho^2
\Im\left(
\int_0^{2\pi}
\hh\psi^\al_+{}'\!\left(\rho e^{i\phi+\Gt}\right)
\hh\psi^\al_-{}'\!\left(\rho e^{i\phi+\Gt}\right) \;
e^{2(i\phi+\Gt)}
d\phi
\right)
\Eqno
$$
where $\hh\psi^\al_\pm:\projal\ME\mapsto\Cmpx$ is defined in \defhhpsi.
Since  ${\bf S}$ has holes in it  this
integral cannot be extended to all of ${\bf S}$.
However these expressions are proportional to
the real and imaginary parts of an integral of an analytic
function:
$$
\ang^{\al} =
8
\Re(\Intr^\al_1)
{\qquad\rm and\qquad}
\eng^{E\al}(\tau) =
\left( {f(\tau)\over h(\tau)} \right)^{1 \over 2}
8
\Im(\Intr^\al_1)
\Eqno\label\reandim
$$
where by substituting $z=\rho e^{i\phi}$
$$
\Intr^\al_1 =
\oint_{|z|=\rho e^{\Gt}}
\hh\psi^\al_+{}'(z)\,
\hh\psi^\al_-{}'(z)\,
z\,
dz + \half\pi i A^\al\cnj{A^\al} \,.
\Eqno\label\defIone
$$
By Cauchy's theorem this integral is invariant under continuous
deformations of the contour  and  is therefore independent of $\tau$ in the
region $\Eal$. Thus we may write in this region
$$
\Intr^\al_1 =
\oint_{|z|=\rho}
\hh\psi^\al_+{}'(z)\,
\hh\psi^\al_-{}'(z)\,
z\,
dz + \half\pi i A^\al\cnj{A^\al}
$$
so
$$
\eng^{E\al}(\tau) =
-4\left( {f(\tau)\over h(\tau)} \right)^{1 \over 2}
\Re\left(2\rho^2
\int_0^{2\pi}
\hh\psi^\al_+{}'\!\left(\rho e^{i\phi}\right)
\hh\psi^\al_-{}'\!\left(\rho e^{i\phi}\right)
e^{2i\phi}
d\phi
+ \pi  A^\al \cnj{A^\al}
\right)
\Eqno
$$
and
$$
\ang^{E\al} =
8\rho^2
\Im\left(
\int_0^{2\pi}
\hh\psi^\al_+{}'\!\left(\rho e^{i\phi}\right)
\hh\psi^\al_-{}'\!\left(\rho e^{i\phi}\right)
e^{2i\phi}
d\phi
\right)\,.
\Eqno\label\angEE
$$
By substituting  \solconver\ into \angEE\ we get \angLL\ and note that
$$
\ang^{E\al}=\ang^{L\al}\mathrel{\mathop=^{\rm def}}\ang^\al \,.
$$

Although $\eng^{E\al}$ has little immediate physical significance we shall
shall refer to it as a {\it pseudo energy} in the following.

\Section{A Conservation Identity}

Given any solution in the asymptotic Lorentzian
domain of any cylinder it is now
possible to use the previous
theorem  to construct a global solution compatible with
our continuity and junction conditions.
 This is the analogue of solving a Cauchy
problem for the field on $M$.
One can then compute the energy and
 momentum currents in the asymptotic Lorentzian
domain of any other cylinder. In general the energy and  momentum are
not globally conserved. In itself this is not surprising
 since the field $\psi$
propagates through  gravitational fields in addition to being  diffracted
through Euclidean domains.
By axial symmetry the momentum is always conserved for a topology with a pair
of collinear cylinders. We shall also demonstrate below  that for
topologies containing any number of cylinders, two of which are collinear,
solutions, monochromatic in one of these collinear cylinders, give rise to
momentum  conservation.
No similar statements are possible for the energy.
However the canonical
expressions for the energy and momentum forms are inspired by symmetry
considerations that ensure the forms constructed with the aid of Killing
vector fields are locally closed and hence
locally conserved. In the presence of metric degeneracy such arguments
break down and it is of interest to search for alternative quantities that
may be conserved in the  global topology under consideration.

Motivated by the pseudo-energy and momentum integrals above define
$$
\Intr^\al_n =
\oint_{\projal(\Sigal)}
\tt\psi^\al_+{}'(z)\,
\tt\psi^\al_-{}'(z)\,
z^n\,
dz
$$
where
$$
\eqalign{
\tt\psi^\al_\pm{}' &: \projal(\ME) \mapsto\Cmpx \cr
\tt\psi^\al_+{}'(z) &= \psi^\al_+(z) +
\sum_{\cendel\ne\oppal} \half A^\del ( z-\prohal(\cendel))^{-1} \cr
\tt\psi^\al_-{}'(z) &= \psi^\al_+(z) +
\sum_{\cendel\ne\oppal} \half \cnj{A^\del} (z-\prohal(\cendel))^{-1} \cr
}
$$
are well defined functions.

Also let
$$
\Ang^\al=
(-R_1 \Intr^\al_0 + \oneoverfourR \Intr^\al_2 ) e^\al_1 -
i(R_1 \Intr^\al_0 + \oneoverfourR \Intr^\al_2) e^\al_2 +
\Intr^\al_1 e^\al_3
\in\Cmpx^3
$$
where $\{e^\al_1,e^\al_2,e^\al_3\}$ is a natural orthonormal basis
 for $\Real^3$.
We embed  $M$ in  $\Real^3$ with the axis of  cylinder $\lal$  pointing
in the direction of $e^\al_3$.
The projection of $\Ang^\al$ along $e^\al_3$ is $\Intr^\al_1$
which agrees with 
definition \defIone\
since
$$
\eqalign{
&
\oint_{\projal(\Sigal)}
\tt\psi^\al_+{}'(z)\,
\tt\psi^\al_-{}'(z)\,
z\,
dz
 =
\oint_{\projal(\Sigal)}
\left(\hh\psi^\al_+{}'(z)+\half A^\al z^{-1}\right)
\left(\hh\psi^\al_-{}'(z)+\half \cnj{A^\al} z^{-1}\right)
z\,
dz \cr
& \qquad=
\oint_{\projal(\Sigal)}
\hh\psi^\al_+{}'(z)\,
\hh\psi^\al_-{}'(z)\,
z\,
dz
+
\half\pi i A^\al \cnj{A^\al}
+
\half\oint_{\projal(\Sigal)}
\left(
\cnj{A^\al}\hh\psi^\al_+{}'(z)+
A^\al\hh\psi^\al_-{}'(z)
\right) dz
}
$$
and the last term is zero because $\psi^\al_\pm{}'$ has no term in $z^{-1}$.
The real and imaginary parts of $\Intr^\al_1$   determine the
momentum and pseudo-energy by \reandim.

We assert:

{\bf Theorem 2}
$$
\sum_\al \Ang^\al=0 \,.
\Eqno\label\lsum
$$
{\bf Proof}
\newline
A change of chart $\lal\to\lbe$ induces
$z\to y=M^{-1}(z)=\projbe\projal{}^{-1}(z)$
hence
$$
\tt\psi_\pm^\be{}'(z)
=
{d \tt\psi_\pm^\be(z) \over dz}
=
{d \over dz} (\tt\psi_\pm^\al(y))
=
\tt\psi_\pm^\al{}'(y){dy \over dz}
=
{\tt\psi_\pm^\al{}'(y) \over M'(y)} \,.
$$
If $C:\Real\mapsto\Cmpx$ is any closed curve then
$$
\oint_{\projbe(C)}
\tt\psi^\be_+{}'(z)
\tt\psi^\be_-{}'(z)
z\,
dz
=
\oint_{\projal(C)}
\tt\psi^\be_+{}'(y)
\tt\psi^\be_-{}'(y)
{M(y) \over M'(y)}
dy\,.
$$
Let $\{a_1^\albe,a_2^\albe,a_3^\albe \}$ be the Euler angles introduced
above
so that  using \projdef\
$$
{M(y) \over M'(y)}
=
y \cos(a_2^\albe)
+
\sin(a_2^\albe)
\left(-R_1 e^{-i a_1^\albe} + {\oneoverfourR y^2  e^{i a_1^\albe}} \right) \,.
$$

By a change of  basis from $\{e^\al_1,e^\al_2,e^\al_3\}$ to
$\{e^\be_1,e^\be_2,e^\be_3\}$
{\tenpoint
$$
\eqalign{
&\pmatrix{
e^\al_1 & e^\al_2 & e^\al_3
}
= \cr
&
\pmatrix{
e^\be_1 & e^\be_2 & e^\be_3
}\!\!
\pmatrix{
\cos(\!\aabn3\!) & \sin(\!\aabn3\!) & 0 \cr
\sin(\!\aabn3\!) & \cos(\!\aabn3\!) & 0 \cr
0            &      0       & 1
}\!\!
\pmatrix{
\cos(\!\aabn2\!) & 0 & \sin(\!\aabn2\!) \cr
0            & 1 &      0       \cr
\sin(\!\aabn2\!) & 0 & \cos(\!\aabn2\!)
}\!\!
\pmatrix{
\cos(\!\aabn1\!) & \sin(\!\aabn1\!) & 0 \cr
\sin(\!\aabn1\!) & \cos(\!\aabn1\!) & 0 \cr
0            &      0       & 1 }
}
$$
}
giving
$$
e^\be_3 \cdot
\pmatrix{
e^\be_1 & e^\be_2 & e^\be_3
}
=
\pmatrix{
\sin(\aabn2)\cos(\aabn1) & \sin(\aabn2)\sin(\aabn1)
& \cos(\aabn2)
}\,.
$$
Now if $f:\projal(M^\E)\mapsto\Cmpx$ is any analytic function then
$$
\sum_\al \oint_{\projbe(\Sigal)} f(z)\,dz = 0\,.\Eqno
$$
So with $f(z)=\tt\psi_+^\be{}'(z)\, \tt\psi_-^\be{}'(z)\, z$
$$
\eqalignno{
0 &=
\sum_\al \oint_{\projbe(\Sigal)}
\tt\psi_+^\be{}'(z)\, \tt\psi_-^\be{}'(z)\, z\,dz
\cr & =
\sum_\al \oint_{\projal(\Sigal)}
\tt\psi_+^\al{}'(z)\, \tt\psi_-^\al{}'(z)\,
{M(z) \over M'(z)}
\,dz
\cr & =
\sum_\al \oint_{\projal(\Sigal)}
\tt\psi_+^\al{}'(z)\, \tt\psi_-^\al{}'(z)\!\!
\left(z \cos(a_2^\albe)
+
\sin(a_2^\albe)
\left(-R_1 e^{-i a_1^\albe} + {\oneoverfourR z^2  e^{i a_1^\albe}} \right)
\!\right)\! dz
\cr & =
\sum_\al
\Intr^\al_1 \cos(a_2^\albe)
+
\sin(a_2^\albe)
\left(
-R_1 e^{-i a_1^\albe}
\Intr^\al_0
+
{\oneoverfourR e^{i a_1^\albe} }
\Intr^\al_2
\right)
\cr & =
\sum_\al
\Intr^\al_1 \cos(a_2^\albe)
+
\sin(a_2^\albe)\cos(a_1^\albe)
(-R_1 \Intr^\al_0 + \oneoverfourR \Intr^\al_2 )
\cr
&\qquad\qquad
-
\sin(a_2^\albe)\sin(a_1^\albe)
i(R_1 \Intr^\al_0 + \oneoverfourR \Intr^\al_2)
\cr & =
\sum_\al
\sin(a_2^\albe)\cos(a_1^\albe)
(\Ang^\al \cdot e^\al_1)
+
\sin(a_2^\albe)\sin(a_1^\albe)
(\Ang^\al \cdot e^\al_2)
+
\cos(a_2^\albe)
(\Ang^\al \cdot e^\al_3)
\cr & =
\sum_\al
(\Ang^\al \cdot e^\al_1)
(e^\al_1 \cdot e^\be_3)
+
(\Ang^\al \cdot e^\al_2)
(e^\al_2 \cdot e^\be_3)
+
(\Ang^\al \cdot e^\al_2)
(e^\al_2 \cdot e^\be_3)
\cr & =
\left(\sum_\al
\Ang^\al \right)\cdot e^\be_3
\qquad\qquad\forall\lbe
\,.}
$$
If the set $\{e^\be_3\}$ span $\Real^3$ then this implies $\sum_\al
\Ang^\al=0$~.
In the case where the $\{e^\be_3\}$ span  $\Real^2$ or $\Real^1$
we compare with a  topology containing an additional cylinder $\gamma$
transverse to the existing set. By Cauchy's theorem
$\Intr_n^\gamma=0$ for $n=0,1,2$ so
$\Ang^\gamma =0$.
The expression
$\left(\sum_\al
\Ang^\al \right)\cdot e^\be_3=0
$
above  becomes
$\left(\sum_\al
\Ang^\al \right)\cdot e^\gamma_3
+
\Ang^\gamma \cdot e^\gamma_3=0
$
where now $\{e^\gamma_3\}$ span $\Real^3$. \qed

An immediate corollary is that if
 $M$ has one cylinder only it cannot  admit a regular
solution with non-zero momentum.

{\bf Proof:} $\Ang^\al=0$ hence $\Intr_1^\al=0$ and $\ang^\al=0$. \qed

We can  also calculate $\Intr^\al_n$ in the chart $\Phil(x)=(\tau,\phi)$
using the solutions $\psi^{L\al}_\pm$
 on the Lorentzian
region $\Lal\union\Mal$ restricted to $\Sigal$:
$$
\eqalign{
\Intr^\al_n =
{1 \over 4} (\rho_\al)^{n-1} \int_0^{2\pi}
e^{i\phi(n-1)}
\Big( &
2\psi^{L\al}_-{}'(\phi)\cnj{\psi^{L\al}_-{}'(\phi)}
-
2\psi^{L\al}_+{}'(\phi)\cnj{\psi^{L\al}_+{}'(\phi)}
\cr
& -
2i\psi^{L\al}_+{}'(\phi)\cnj{\psi^{L\al}_+{}'(\phi)}
-
2i\psi^{L\al}_+{}'(\phi)\cnj{\psi^{L\al}_+{}'(\phi)}
\cr
& +
\cnj{A^\al}\left(
(1-i)\psi^{L\al}_+{}'(\phi) + (1+i)\psi^{L\al}_+{}'(\phi)
\right) +
\cr
&
A^\al\left(
(1-i)\cnj{\psi^{L\al}_+{}'(\phi)} + (1+i)\cnj{\psi^{L\al}_+{}'(\phi)}
\right)
+ i A^\al\cnj{A^\al}
\Big) d\phi
\,.}
$$
Thus the sum-rule \lsum\  can be seen to correlate properties of the Lorentzian
solutions.

\Section{Constraints on the Manifold for the
Propagation of Monochromatic Modes}

In the introduction we alluded to the general
fact that proving the existence of regular
solutions to tensor equations
on a manifold is a problem in global analysis.
In this section we construct a particular global solution  corresponding
to a monochromatic, positive energy, propagating mode in
a preferred cylinder. The nature of this construction will clearly indicate
what topological constraints must be imposed for it to exist on $M$. A space
of solutions of this type is relevant to the construction of the Fock  spaces
that feature in a  field quantisation. The classical solutions below may be
used to construct the Bogoloubov coefficients necessary to estimate
particle production induced by a degenerate gravitational field.

Let
the cylinder $\beta$
be oriented at an angle $\aabn2=\omega$ with
the cylinder $\alpha$.
In constructing the map between charts adapted to these cylinders  we choose
the Euler angles  $\{ \aabn1,\aabn2,\aabn3\}$ with $\aabn1=0$ and $\aabn3=0$.
The required Mobius transformation is then given by:
$$
\projbe\projal\tmone(z)
=
{ -2R_1\tan\omegabytwo + z
\over\displaystyle
1 + {\strut \tan\omegabytwo z \over 2R_1 }
}\,.
\Eqno
$$
Introduce  the metric  constants $\xi_\al$ and $\xi_\be$ by the equations
$\rho_\al=2R_1\tan(\xi_\al/2)$, $\rho_\be=2R_1\tan(\xi_\be/2)$
where $\rho_\al$ and $\rho_be$ are defined by \defrho . Also the constants
$\epsilon^\al$,  $\epsilon^\be$ follow from \defeps.
On $(\tau,\phi)\in\Phil(\LuM)$ define the null
coordinates
$$
\gamma_\pm^\al = \phi \pm G^\al(\tau)
\qquad\hbox{such that}\qquad
\gamma_\pm^\al = \phi \pm {\tau -\epsilon^\al \over R^\al_2}
{\rm \ \ on\ }\Mal
\Eqno
$$
and similarly for $\lbe$.
For $x\in\LuM$ with
$(\tau,\phi)=\Phil(x)$
let the solution to the wave equation  be the left-moving single mode with
frequency $k\in\Intg^+$:
$$
\psi^\al|_\LuM(x)
=
e^{i k\gamma^\al_+} \,.
$$
Therefore from \solconver\ ,
$\psi_+^{\L\al}(\gamma^\al_+)=e^{i k\gamma^\al_+}$ and
$\psi_+^{\L\al}(\gamma^\al_-)=0$
so
$$
\eqalign{
\psi_+^{\L\al}(\gamma)
&=
\half (1+i)
\left(
\psi_+^{\al}(\rho_\al e^{i\gamma}) -
i\;\cnj{\psi_-^{\al}(\rho_\al e^{i\gamma})}
\right)
=
e^{ik\gamma}
\cr
\psi_-^{\L\al}(\gamma)
&=
\half (1-i)
\left(
\psi_+^{\al}(\rho_\al e^{i\gamma}) +
i\;\cnj{\psi_-^{\al}(\rho_\al e^{i\gamma})}
\right)
= 0
\,. }
\Eqno
$$
Solving these we get
$$
\eqalign{
\psi_+^{\al}(\rho_\al e^{i\gamma})
&=
\half(1-i)e^{ik\gamma}
\cr
\psi_-^{\al}(\rho_\al e^{i\gamma})
&=
\half(1-i)e^{-ik\gamma}
\,. }
$$
Analytically continuing these to the domain $\projal(\ME)$ yields
$$
\eqalign{
\psi_+^{\al}(z)
&=
\half(1-i)\rho_\al^{-k} z^k
\cr
\psi_-^{\al}(z)
&=
\half(1-i)\rho_\al^k {z^{-k}}
\,. }
\Eqno\label\onefE
$$
Hence there are no log terms for this particular solution. Furthermore
since  $\psi^\al_+\sim z^k$ then $\oppal\not\in\Proj(\ME)$.
In
order that this solution be regular we have excised the point $z=\infty$.
This is equivalent to requiring that a cylinder $\beta$
exist such that $\Proj (\Sigbe)$  enclose $\oppal$.

In order to describe the above solution on any
 $\lbe$ cylinder we must transform
it to $(\Ube,\Phibe)$ coordinates.
{}From \conME\ we have
$$
\eqalign{
\psi_+^\be(z)
&=
\psi_+^\al\circ \projbe\projal\tmone(z)
=
\half (1-i)
\left({
-\tan(\half\omega) + z
  \over
\tan(\half\xi_\al) ( 1 +\tan(\half\omega)z )
}\right)^k
\cr
\psi_-^\be(z)
&=
\psi_-^\al\circ \projbe\projal\tmone(z)
=
\half (1-i)
\left({
\tan(\half\xi_\al) ( 1 +\tan(\half\omega)z )
  \over
-\tan(\half\omega) + z
}\right)^k
\,. }
$$
Thus from \solconver, with $z=\rho_\be e^{i\gamma^\be}$ ,
$(\tau,\phi)=\Phibe(x)$
$$
\eqalign{
&\psi|_\LuMbe(x)
=
\psi_+^{\L\be}(\gamma^\be_+) +
\psi_-^{\L\be}(\gamma^\be_-) \cr
& \quad =
\half (1+i)
\left(
\psi_+^{\be}(\rho_\be e^{i\gamma^\be_+}) -
i\;\cnj{\psi_-^{\be}(\rho_\be e^{i\gamma^\be_+})}
\right)
+
\half (1-i)
\left(
\psi_+^{\be}(\rho_\be e^{i\gamma^\be_-}) +
i\;\cnj{\psi_-^{\be}(\rho_\be e^{i\gamma^\be_-})}
\right) \cr\cr
& =
\half\!
\left({
-\tan(\half\omega) + \tan(\half\xi_\be) e^{i\gamma^\be_+}
  \over
\tan(\half\xi_\al) ( 1 +\tan(\half\omega)\tan(\half\xi_\be) e^{i\gamma^\be_+} )
}\right)^k
\!\!\!+\!
\half\!
\left({
\tan(\half\xi_\al) ( 1 +\tan(\half\omega)\tan(\half\xi_\be)
e^{-i\gamma^\be_+} )
  \over
-\tan(\half\omega) + \tan(\half\xi_\be) e^{-i\gamma^\be_+}
}\right)^k
\cr\cr
& -
\ibytwo\!
\left({
-\tan(\half\omega) + \tan(\half\xi_\be) e^{i\gamma^\be_-}
  \over
\tan(\half\xi_\al) ( 1 +\tan(\half\omega)\tan(\half\xi_\be) e^{i\gamma^\be_-} )
}\right)^k
\!\!\!+\!
\ibytwo\!
\left({
\tan(\half\xi_\al) ( 1 +\tan(\half\omega)\tan(\half\xi_\be)
e^{-i\gamma^\be_-} )
  \over
-\tan(\half\omega) + \tan(\half\xi_\be) e^{-i\gamma^\be_-}
}\right)^k \!.
}
\Eqno\label\onefEb
$$
We have seen that for this solution to exist the topology must contain at
least one cylinder attached to the cap containing the point $\oppal$.
(It is interesting to note that this
constraint can be relaxed for standing wave solutions in an
asymptotically flat Lorentzian domain.)
We may use { theorem 2} to calculate the momentum associated with the
solution on this cylinder.
Since $\Intr^\al_1=-8k^2\pi $ , $\Intr^\al_0=0$ and $\Intr^\al_2=0$
then $\Ang^\al=-8k^2\pi  e^\al_3$. Hence
 $\Intr^\be_1=\Ang^\be\cdot e^\be_3=(-\Ang^\al)\cdot e^\be_3=
8k^2\pi  e^\al_3\cdot e^\be_3=-8k^2\pi  \cos(\omega)$
and the momentum in cylinder $\beta$ is  $\ang^\be = 8k^2\pi\cos(\omega)$.
As $\ang^\al=-8k^2\pi$ the momentum is conserved for $\omega=\pi$,
i.e. collinear cylinders.

 The
momentum associated with this solution in any  cylinder other than these
two is zero by Cauchy's theorem.

\def\khalf{{(-1)^k\over 2}}
\def\kibytwo{{i(-1)^k\over 2}}

For collinear cylinders,  $\omega=\pi$, the solution becomes, with
 $(\tau,\phi)=\Phibe(x)$:
$$
\eqalign{
\psi_\LuMbe(x) = &
\khalf  e^{-ik\gamma_+^\be}
\left( (\tan(\half\xi_\al)\tan(\half\xi_\be))^{k} +
  (\tan(\half\xi_\al)\tan(\half\xi_\be))^{-k} \right)
\cr &+
\kibytwo e^{-ik\gamma_-^\be}
\left( (\tan(\half\xi_\al)\tan(\half\xi_\be))^{k} -
  (\tan(\half\xi_\al)\tan(\half\xi_\be))^{-k} \right)
\,.}\Eqno
$$
We can only remove the right-moving wave term, $e^{-ik\gamma_-^\be}$, if
$\tan(\half\xi_\al)\tan(\half\xi_\be)=\pm1$, i.e.
\break\hbox{$\xi_\al+\xi_\be=\pi$,} which
means that the $\Sigma$-rings coincide and there is no Euclidean region.

The solution $\psi^{\L\be}$ diverges when
either $\omega=\xi_\be$ or  $\omega+\xi_\be=\pi $~. The first of these
implies  that $\projbe\Sigbe$  contains
$\cenal$ and  the second implies that $\projbe\Sigbe$ contains $\oppal$.
Since the cylinders have non-zero radii these cannot occur.

The energy of the solution \onefEb\  can be computed using \engLL.
For $k=1$ one finds:
$$
\eqalign{
\eng^{L\be}(\tau)
= &
4\pi\sin^2(\xi_\beta)
\left({-f(\tau) \over h(\tau)}\right)^\half
\cr &
\left(
{ ( 1 + \cos(\omega) \cos(\xi_\beta) )
\over
| \cos(\omega) + \cos(\xi_\beta) |^3 }
{ ( 1 + \cos(\xi_\alpha) ) \over ( 1 - \cos(\xi_\alpha) ) }
+
{ ( 1 - \cos(\omega) \cos(\xi_\beta) )
\over
| \cos(\omega) - \cos(\xi_\beta) |^3 }
{ ( 1 - \cos(\xi_\alpha) ) \over ( 1 + \cos(\xi_\alpha) ) }
\right)
}
\Eqno
$$
which is manifestly positive definite.
\def\ss#1{\!#1\!}
\def\tl{\widetilde}
\Section{Constraints  for the Existence of Solutions
 on a Compact Manifold }

Here we construct regular  solutions in   the  simplest
compact manifold with an  axially symmetric degenerate metric.
It may be visualised as a pair of Euclidean spheres connected by a single
cylinder. See the {\it bone} {\it Figure 5}.
 The metric becomes degenerate on a pair of rings in the cylinder
and partitions its geometry so that the middle section is Lorentzian.
Such a manifold may be viewed as a  simple model of a signature changing
closed cosmology.

{\bf Theorem 3}

The  Lorentzian region of $M$ will admit monochromatic standing wave
solutions to \waveqn\ of frequency $k\in\Intg$ if and only if $\epsilon$
satisfies the condition:
$$
\cos(k\epsilon)=0 \,.\Eqno  $$\label\kmode
where $\epsilon$ is a real parameter determined
by the metric (equation
\def\defneps{({\advance \EQNO by 1 \the\EQNO})}\defneps\
below).

{\bf Proof}\hfill\break
Let $M$ have the $S^2$ topology above where  $\Sigma^1$, $\Sigma^2$ denote
 two non-intersecting rings where the metric changes signature.
Thus $M$ is
$ E^1\union \Sigma^1 \union L\union \Sigma^2\union  E^2 $ where $E^\al$ are the
Euclidean domains and $L$ is the Lorentzian domain. Let $(U^\al , \Phil)$
be charts for $U^\al = E^\al\union \Sigal\union L$ where
$$
\eqalign{
\Phi_\alpha &: U^\al\mapsto\Real^2 \cr
&:x \mapsto(\tau_\al,\phi_\al).
}
\qquad\alpha=1,2.
$$
We note that the $\tau$ coordinates induce opposite coordinate
time orientations on
$L$.
The differential structure $(\Phi_1)\circ(\Phi_2)^{-1}:
(\tau_2,\phi_2)\mapsto (\tau_1,\phi_1)$ is given by the equations:
$$
\eqalign{
\tau_1 + \tau_2 &= \tau_0 \cr
\phi_1 + \phi_2 &= \phi_0 \,.
}$$
This differential structure is compatible with $M$ being orientable.

It is now convenient to write
$$\eqalign{
G^\al &: \tau_\al(\Phi_\al(U^\al))\mapsto\Real \cr
&:\tau_\al\mapsto\int_{\tau_\al(\Sigma^\al)}^{\tau_\al}
\left|{-f^1(\tau') \over h^1(\tau')}\right|^\half
d\tau'
}
$$
where
$\{(\tau_\al,\phi_\al):\tau_\al=\tau_\al(\Sigma^\be)\}=\Phil(\Sigma^\be)$.
For $x\in U^1 \cap U^2 $
$$
\eqalign{
G^1(\tau_1(x)) + G^2(\tau_2(x))
&=
\int_{\tau_1(\Sigma^1)}^{\tau_1(x)}
\left({-f( t_1) \over h( t_1)}\right)^\half
d t_1
+
\int_{\tau_2(\Sigma^2)}^{\tau_2(x)}
\left({-f( t_2) \over h( t_2)}\right)^\half
d t_2
\cr
&=
\int_{\tau_1(\Sigma^1)}^{\tau_1(x)}
\left({-f( t_1) \over h( t_1)}\right)^\half
d t_1
-
\int_{\tau_1(\Sigma^2)}^{\tau_1(x)}
\left({-f( t_1) \over h( t_1)}\right)^\half
d t_1
\cr
&=
\int_{\tau_1(\Sigma^1)}^{\tau_1(\Sigma^2)}
\left({-f( t_1) \over h( t_1)}\right)^\half
d t_1
\cr
&=\epsilon < 0
}\Eqno\label\defneps
$$
where   $t_2=\tau_0- t_1$.
{}From {\bf theorem (1.2)} the solution $\psi|_\L$ in the Lorentzian region
   is given as
$$
\eqalign{
\psi|_\L(x) &=
\psi^{L1}_+(\phi_1\ss+G^1(\tau_1))
+
\psi^{L1}_-(\phi_1\ss-G^1(\tau_1))
=
\psi^{L2}_+(\phi_2\ss+G^2(\tau_2))
+
\psi^{L2}_-(\phi_2\ss-G^2(\tau_2)) \cr
&=
\psi^{L2}_+(\phi_0+\epsilon-\phi_1\ss-G^2(\tau_1))
+
\psi^{L2}_-(\phi_0-\epsilon-\phi_1\ss+G^2(\tau_1))
\,.}
$$
The absence of any log terms is demanded by regularity.

Since $\phi_1\ss+G^1(\tau_1)$ and $\phi_1\ss-G^1(\tau_1)$ are independent
$$
\eqalign{
\psi^{L1}_+(\phi) &=
\psi^{L2}_+(\phi_0+\epsilon-\phi)
\cr
\psi^{L1}_-(\phi) &=
\psi^{L2}_-(\phi_0-\epsilon-\phi)
}\qquad\forall\phi\,.
$$

The Euclidean solutions in $E^1$ and $E^2$ will be constrained to match
this Lorentzian solution.
{}From theorem 1.3 the Euclidean solutions on  $\Sigma^1$ and $\Sigma^2$
 satisfy:
$$
\eqalign{
\psi^2_+(\rho_2 e^{i\phi})=
\half \psi^1_+(\rho_1 e^{i(\phi_0+\epsilon-\phi)})
+ &
\half \psi^1_+(\rho_1 e^{i(\phi_0-\epsilon-\phi)})
+
\cr &
\ibytwo \cnj{\psi^1_-(\rho_1 e^{i(\phi_0-\epsilon-\phi)})}
-
\ibytwo \cnj{\psi^1_-(\rho_1 e^{i(\phi_0+\epsilon-\phi)})}
\cr
\cnj{\psi^2_-(\rho_2 e^{i\phi})} =
\ibytwo \psi^1_+(\rho_1 e^{i(\phi_0+\epsilon-\phi)})
- &
\ibytwo \psi^1_+(\rho_1 e^{i(\phi_0-\epsilon-\phi)})
+
\cr &
\half \cnj{\psi^1_-(\rho_1 e^{i(\phi_0-\epsilon-\phi)})}
+
\half \cnj{\psi^1_-(\rho_1 e^{i(\phi_0+\epsilon-\phi)})}
\,.}
\Eqno\label\eqnpf
$$
It is convenient to write  $\cnj{\psi^\al_\pm(\cnj z)}$  as
$\tl\psi^\al_\pm(z)$. (Observe $\tl\psi^\al_\pm$ are analytic
in $z$.)

The analytic continuation of \eqnpf\ may be written:
$$
\eqalign{
\psi^2_+(z) =
\half \psi^1_+ \!\!
\left({\rho_1 \rho_2 e^{i(\phi_0-\epsilon)} \over z } \right)
+
\half \psi^1_+ \!\!
\left({\rho_1 \rho_2 e^{i(\phi_0+\epsilon)} \over z } \right)
+
\ibytwo \tl\psi^1_-\!\!
\left({\rho_1 z\over \rho_2 e^{i(\phi_0-\epsilon)} } \right)
-
\ibytwo \tl\psi^1_-\!\!
\left({\rho_1 z\over \rho_2 e^{i(\phi_0+\epsilon)} } \right)
\cr
\psi^2_-(z) =
\half \psi^1_- \!\!
\left({\rho_1 \rho_2 e^{i(\phi_0-\epsilon)} \over z } \right)
+
\half \psi^1_- \!\!
\left({\rho_1 \rho_2 e^{i(\phi_0+\epsilon)} \over z } \right)
+
\ibytwo \tl\psi^1_+\!\!
\left({\rho_1 z\over \rho_2 e^{i(\phi_0+\epsilon)} } \right)
-
\ibytwo \tl\psi^1_+\!\!
\left({\rho_1 z\over \rho_2 e^{i(\phi_0-\epsilon)} } \right)
.}
\Eqno$$\label\coneqn
We now express the Euclidean solutions as the general Laurent expansions
$$
\psi^1_+(z)=\sum_{k\le 0}A^1_k z^k
\quad
\psi^1_-(z)=\sum_{k\le 0}B^1_k z^k
\quad
\psi^2_+(z)=\sum_{k\le 0}A^2_k z^k
\quad
\psi^2_-(z)=\sum_{k\le 0}B^2_k z^k
\,.\Eqno\label\eqnpff
$$
The integers $k\le 0$ since $\psi^\al_\pm$ must be bounded as $z\to\infty$.
The functions $\psi^\al_\pm$ are not required to
be bounded as $z\to 0$, since $\projal\S^\al\subset\Cmpx=\Cmpx-D$ where $D$
is a disc at the origin.
Substituting these expansions into
\coneqn\ yields
$$
\eqalign{
\sum_{k\leq 0} A^2_k z^k =
-\sum_{k\leq 0}\cnj{B^1_k} {\rho_1^{\;k} \over\rho_2^{\;k}} z^k e^{-i k \phi_0}
\sin(k\epsilon)
+
\sum_{k\leq 0} A^1_k {\rho_1^{\;k}  \rho_2^{\;k} \over z^k} e^{i k \phi_0}
\cos(k\epsilon)
\cr
\sum_{k\leq 0} B^2_k z^k =
\sum_{k\leq 0} \cnj{A^1_k} {\rho_1^{\;k} \over\rho_2^{\;k}} z^k e^{-i k \phi_0}
\sin(k\epsilon)
+
\sum_{k\leq 0} B^1_k {\rho_1^{\;k}  \rho_2^{\;k} \over z^k} e^{i k \phi_0}
\cos(k\epsilon)
\,.}
$$
Hence comparing coefficients of $z$ we deduce $\cos(k\epsilon)=0$
and
$$
A^2_k = -\cnj{B^1_k} {\rho_1^{\;k} \over \rho_2^{\;k}} e^{-i k \phi_0}
\sin(k\epsilon)
\quad{\rm and}\quad
B^2_k =
\cnj{A^1_k} {\rho_1^{\;k} \over \rho_2^{\;k}} e^{-i k \phi_0}
\sin(k\epsilon) \quad\qed
$$

For a given $\epsilon$ we may now write the  solution in the Lorentzian
region as
$$
\eqalign{
 \psi|_\L((\Phi^1){}^{-1}(\tau,\phi))=
\sum_{k\leq 0}\rho_1^{\;k}
\left( A^1_k e^{ik\phi} + B^1_k e^{-ik\phi} \right)
\left( \cos(k G^1(\tau)) - \sin(k G^1(\tau)) \right)
}\Eqno\label\bonesol
$$
where for convergence
$$
\sum_{k\leq 0}
\rho_\al^{\;k} \left( |A^\al_k| + |B^\al_k| \right)
< \infty
\qquad \lal=1,2\,,
$$
and it is understood that  only those modes satisfying \kmode\ are to
be included in the sums above.
We note that \kmode \ admits no (non-constant) solution for $\epsilon =0$. This
is consistent with Louiville's theorem.

Also for \kmode\ to admit solutions $\epsilon/\pi$ must be rational and
$$
{\epsilon \over \pi} = { 2m + 1 \over 2 k } \qquad
\hbox{ where } m,k\in\Intg \,.
$$
Furthermore,  given $\epsilon$, if $k$ is a solution to \kmode\ so is $(2m+1)k$
for all
$m\in\Intg$.
{}From the corollary to theorem 2 it follows that all solutions for this
Manifold have zero
 momentum, $\ang=0$. By direct computation, the energy  associated with
\bonesol\  is:
$$
\eng^L(\tau)
=
\left({-f^1(\tau) \over h^1(\tau)}\right)^\half
\sum_{k\leq 0}  {4\pi k^2 \rho_1^{\;2k}}(|A^1_k|^2 +|B^1_k|^2 )
$$
where the summation is restricted as above.

\Section{Discussion}

We have analysed the massless wave equation on a class of two dimensional
manifolds with smooth degenerate metrics. We have
drawn particular attention to the interplay between the topological
structure of the manifold and the  existence
of regular solutions. Such solutions are
piecewise smooth but globally $C^1$. The difference between  solutions
on compact and non-compact manifolds has been stressed and the effects of
the topology on the currents induced by the Killing symmetries of the metric
have been explicitly calculated. We have also introduced a sum rule that
may be interpreted as a conservation law for a new type of momentum in the
presence of signature change.

Topology change in two dimensional field theories is relevant in a number
of contemporary problems. String field theory interactions proceed by
processes in which the classical two dimensional world sheet exhibits a
change in topology. Indeed {\it Figure 1} may be regarded as a non-compact
immersion in
a spacetime describing the interaction of a set of closed strings.
In such a description the metric on the world sheet is dynamically induced
by an extremal immersion. However if the immersion is in a spacetime
of Lorentzian signature then the induced metric cannot  be
non-degenerate. In order to accommodate the change in signature of
the induced metric it is necessary to confront the problems discussed in
this paper.

The above results pertain to  two dimensional manifold with genus zero.
However it is straightforward in principle to extend these techniques to
the case of non-zero genus.
One may
also expect that the general features discussed above will have analogues in
higher dimension.  Such results can then be interpreted  as the effects of
spacetime topology on the propagation of matter in the presence of
signature change.
\Ref{G Ellis, A Sumeruk, D Coule, C Hellaby  Class. Q. Grav. 1535  {\bf 9}
1992}
\Ref{T Dereli, \"Onder, R W Tucker, Phys. Lett. {\bf B324} (1994) 143}
\Ref{T Dereli,  R W Tucker, Class. Quantum Grav. {\bf 10} (1993)
365}
\Ref{T Dereli, \"Onder, R W Tucker, Class. Quantum Grav. {\bf 10} (1993)
1425}
\Ref{S S Hayward, Class. Q. Grav. L7 {\bf 10  } 1993}
\Ref{M Kossowski, M Kriele  Class. Q. Grav. 1157 {\bf 10} 1993}
\Ref{M Kossowski, M Kriele  Class. Q. Grav. 2363 {\bf 10} 1993}
\Ref{M Kossowski, M Kriele  Proc. Roy. Sc. Lond. 297 {\bf A444} 1994}
\Ref{R Kerner, J Martin Class. Q. Grav. 2111 {\bf 10} 1993}

 If changes of signature
can occur on a Planckian scale then it is imperative to understand how to
extend standard quantum field theory to such topologically non-trivial
backgrounds. We have surmised elsewhere
\Ref{T Dray, C A Manogue, R W Tucker, Gen. Rel. Grav. {\bf 23} (1991) 967},
\dray \  that asymptotically flat
Lorentzian domains that are connected via a Euclidean domain may induce matter
interactions that can be interpreted as particle creation by analogy with
particle creation by localised gravitational curvature.
The results in this paper are an attempt to provide a rigorous background
for the formulation of such a field theory of topologically induced particle
interactions.

\Section{ Acknowledgments }

This work has benefited from useful discussions with Charles Wang.
J G is grateful to the University of Lancaster for a University Research
Studentship and R W T for
support  from the Human Capital and Mobility Programme of the EC.


\vfill
\eject

\References

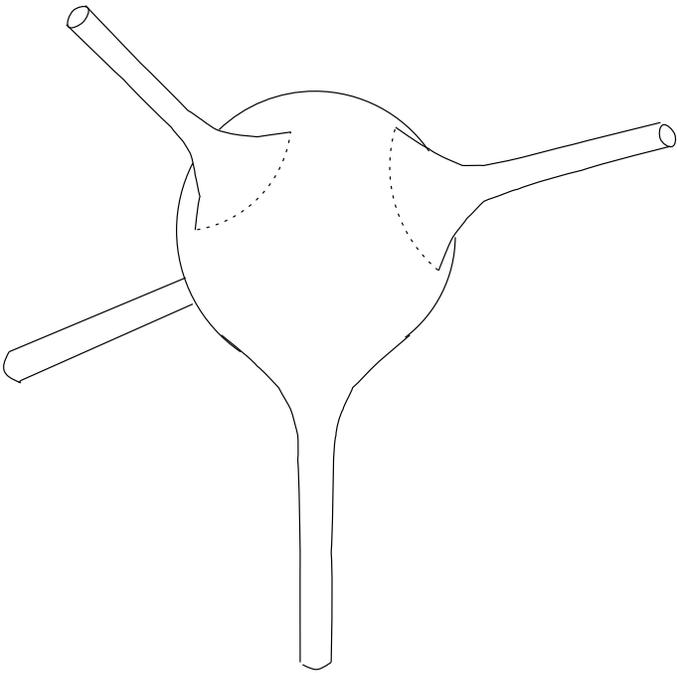
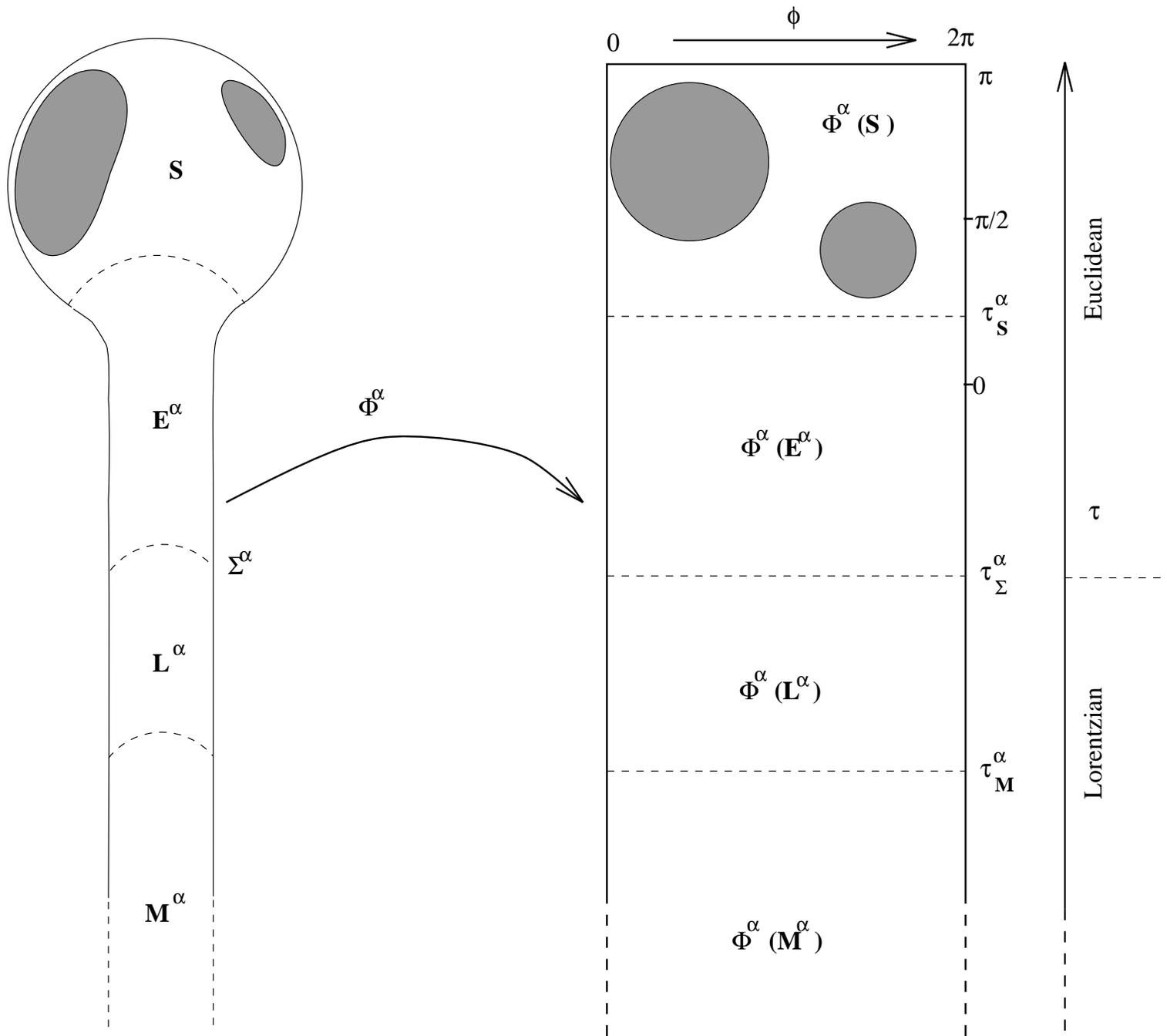
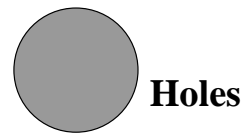
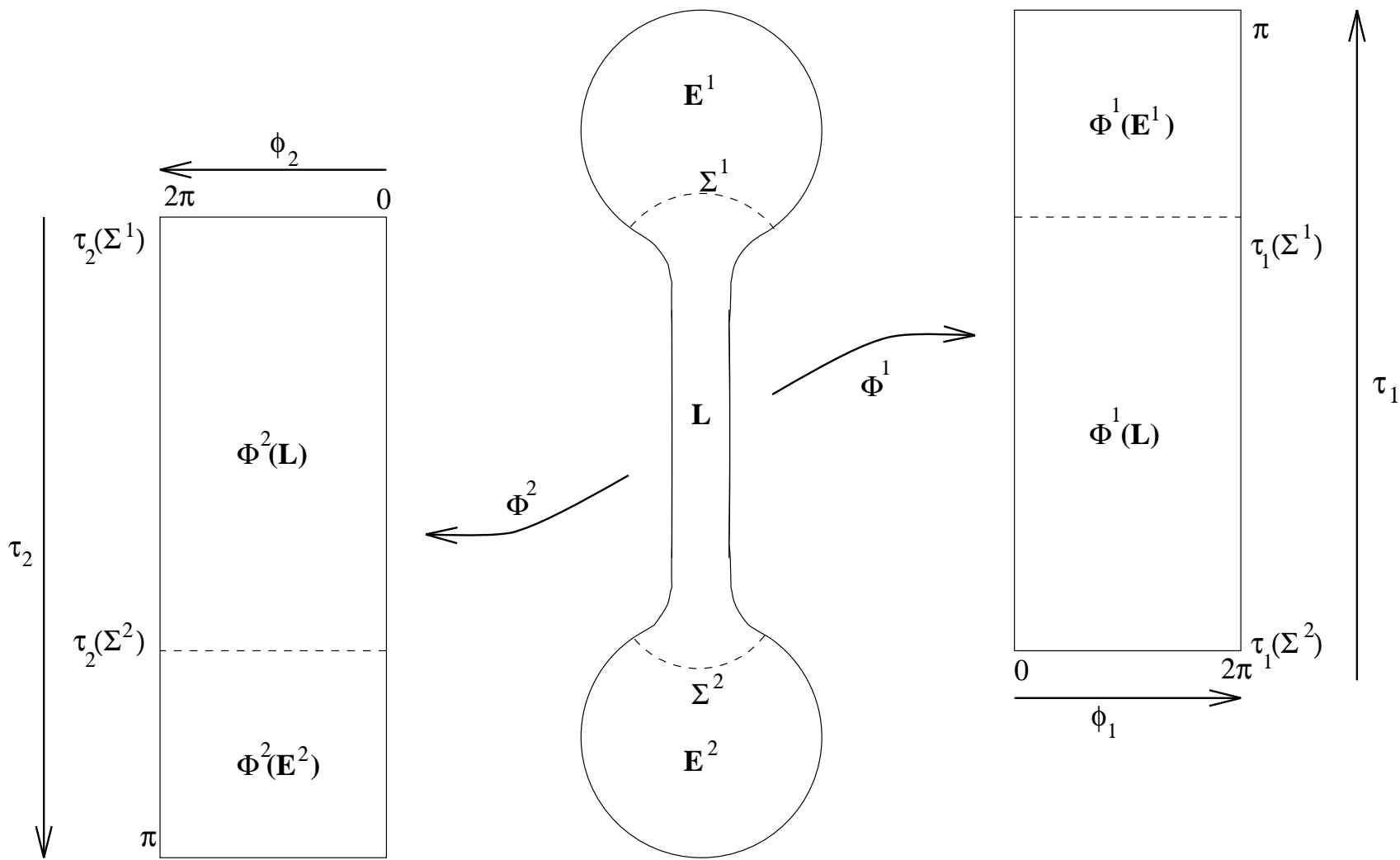
\end